\documentclass[aps,superscriptaddress,twocolumn]{revtex4-2}
\usepackage{amsmath}    
\usepackage{graphicx}   
\usepackage{verbatim}   
\usepackage{color}      
\usepackage{subfigure}  
\usepackage{hyperref}   
\usepackage{bm}
\usepackage{epstopdf}
\graphicspath{{./figures/}}
\usepackage{float}
\usepackage{chngcntr} 
\usepackage{xcolor, soul} 

\usepackage{booktabs,tabularx,makecell}

\usepackage{xcolor}

\usepackage{bibentry}

\newcommand{\ignore}[1]{}
\newcommand{\nobibentry}[1]{{\let\nocite\ignore\bibentry{#1}}}

\newcommand{\bo}[1]{\textbf{#1}}

\begin{document}

\title{Emergent electric fields driven by phonon-coupled skyrmion resonances}

\author{Seno Aji}
\email{senji77@sci.ui.ac.id}
\affiliation{Department of Physics, Faculty of Mathematics and Natural Sciences, Universitas Indonesia, Depok 16424, Indonesia.}


\begin{abstract}
We develop a coarse-grained theoretical description of the macroscopic emergent electric field generated by phonon-coupled lattice deformations in the breathing and rotational dynamics of a skyrmion lattice under microwave excitation. The analysis identifies the symmetry and dynamical conditions that yield rectified (dc) and oscillating (ac) electric fields, even in the absence of net translational motion of the skyrmion lattice, particularly in the dilute-lattice limit. Using experimentally measurable skyrmion profile parameters such as the equilibrium radius, domain-wall width, and dynamical resonance frequency of skyrmion lattice, the model further enables identification of harmonic components contributing to the observed macroscopic electrodynamic response in the long-wavelength phonon limit ($q\rightarrow0$) and at finite phonon frequency, providing a unified framework for phonon-driven spin–charge–lattice coupling in topological magnets.

\end{abstract}
\maketitle

\section{Introduction}
Topologically non-trivial spin textures such as magnetic skyrmions have emerged as a central topic in modern condensed-matter physics owing to their rich electrodynamic properties and potential for spintronic applications \cite{Nagaosa2013_NatNano,Fert2017_NatRevMat,Schulz2012_NatPhys,Rosch2013_NatNano,Li2023_IDM2}. A skyrmion carries an integer topological charge that acts as an emergent magnetic flux for conduction electrons, giving rise to a quantized Berry curvature in real space \cite{Verma2022_SciAdv,Xiao_RevModPhys.82.1959,Zhang_PhysRevB.101.024420}. Furthermore, when conduction electrons flow through a time-dependent skyrmion texture, they acquire additional Berry-phase contributions that serve as an emergent electric field, resulting in spin-motive forces and topological Hall effects \cite{Koide_PhysRevB.100.014408,Tokura2021_ChemRev,Lin2016_PRB,Garst2017_JPhysD,Neubauer2009_PRL,Tatara_PhysRevLett.92.086601,Lucassen_PhysRevB.84.014414,PhysRevB.108.054445}. These emergent electrodynamic phenomena bridge the microscopic spin dynamics of the skyrmion texture with measurable macroscopic transport responses, establishing magnetic skyrmions as model systems for studying topological charge and spin transport \cite{EverschorSitte2018_JAP,Okamura2013_NatCommun}.

Beyond the motion of isolated skyrmions, the collective dynamics of a skyrmion lattice (SkX) exhibit an even richer spectrum of excitation modes, including breathing, clockwise (CW) and counterclockwise (CCW) rotational modes, and coupled translational-phonon excitations \cite{Mochizuki2012_PRL,Iwasaki2013_NatCommun,Mochizuki2014_NatMater}.
Experimentally, microwave absorption and spin-wave spectroscopy have revealed distinct resonance modes associated with these collective degrees of freedom \cite{Onose2012_PRL,Okamura2013_NatCommun}. Theoretically, the SkX can be regarded as an elastic medium endowed with a Berry-phase kinetic term, leading to non-trivial coupling between translational and internal (breathing/rotational) motions \cite{Hu2019_PRB_214412,Lin2015_PRB}.

In the long-wavelength limit, the skyrmion lattice supports emergent phonons whose dispersion remains finite as the wave vector $|q|\rightarrow0$, reflecting the non-Galilean (gyrotropic) nature of the magnetic texture \cite{Petrova2011_PRB,Hu2019_PRB_144424}. While the emergent electromagnetic fields arising from rigid translational motion are well established \cite{Schulz2012_NatPhys,Birch2024}, the coarse-grained electric fields from internal SkX deformations remain much less explored. Experiments show that breathing/rotational resonances are electric-field active~\cite{Okamura2013_NatCommun}, theory predicts both rectified (dc) and ac spin-motive voltages arising from the resonant dynamics of the microwave-driven internal modes of a skyrmion crystal under a tilted magnetic field \cite{Koide_PhysRevB.100.014408}. Concurrently, magnetoelastic coupling-mediated by static strain or propagating acoustic phonons could distort the skyrmion lattice and modify its collective excitation spectrum \cite{Hu2019_PRB_144424, Spachmann2021_PRB_103,Nii2015_NatCommun, Adachi_PhysRevB.109.144413}


At finite frequency, lattice distortions modify the local Dzyaloshinskii–Moriya interaction and exchange stiffness, thereby inducing an inhomogeneous modulation of the skyrmion radius and helicity \cite{Spachmann2021_PRB_103,Masell_PhysRevB.101.214428,Hu_2017,Nomura_PhysRevLett.122.145901}. Such magnetoelastic coupling could drive out-of-phase oscillations between polar and azimuthal spin components, breaking spatial and temporal symmetries necessary for generating a net (dc or low-frequency) emergent electric field. Understanding this phonon-coupled electrodynamics is thus crucial for exploring dynamical spin-charge-lattice interconversion mechanisms in topological magnets.

In this work, we develop a continuum theory based on a coarse-grained description of the skyrmion-lattice dynamics coupled to long-wavelength lattice deformation and phonon drives. Starting from the Berry-phase gauge formalism, {\color{black} we derive the local emergent electric fields in the adiabatic strong-coupling limit of the $sd$-exchange interaction, where the conduction-electron spin follows the slowly varying skyrmion texture 
$\bm{n}(\bm{r},t)$. The skyrmion lattice is treated as a deformable medium whose spin texture responds to long-wavelength lattice strain and phonon excitations. We then perform a coarse-graining over the skyrmion-lattice unit cell to obtain the macroscopic response}. We show that the interplay between the breathing amplitude, rotational phase, and lattice strain field produces non-vanishing spatial averages of $\bm{E}^{e}$ when inversion or rotational symmetry is broken, yielding rectified electric signals even in the absence of net skyrmion translation.
Our results elucidate the microscopic origin and symmetry conditions for phonon-driven emergent electric fields in skyrmion crystals, providing a unified picture of their magnetoelastic electrodynamics.

\section{Emergent Electrodynamics}

The emergent (Berry-phase) electromagnetic fields experienced by a conduction electron whose spin adiabatically follows a smooth, possibly time-dependent magnetization texture $\bm{n}(\bm{r},t)=(\sin \Theta \cos \Phi, \sin \Theta \sin \Phi, \cos \Theta)$ (with $|\bm{n}|=1$), can be derived from a standard $sd$-exchange model \cite{Zhang_PhysRevLett.102.086601, Everschor-Sitte_10.1063/1.4870695},

\begin{align}
    i\hbar\partial_t\psi= \left(\frac{\bm{p}^2}{2m} - J_{sd} \bm{n}(\bm{r},t) \cdot \bm{\sigma} \right)\psi.
\end{align}

\noindent {\color{black} Here $\psi=(\psi_\uparrow,\psi_\downarrow)^{T}$ is the electron spinor, $J_{sd}>0$ is the exchange coupling, and $\sigma$ are the Pauli matrices. In the adiabatic limit, the exchange splitting between spin bands, $\Delta_{sd}=2J_{sd}$, is considerably large compared to the characteristic energy scales associated with the spatial and temporal variations of $\bm{n}$. As a result, the electron spin locally aligns with the magnetization texture. In this regime, the electron spin can be rotated into the local spin frame by a unitary SU(2) transformation, $U(\bm{r},t)=\exp{(-i\frac{\Theta}{2} \bm{\sigma} \cdot \bm{v})}$, which maps the laboratory $z$-axis onto $\bm{n}(\bm{r},t)$, where $\Theta$ is the rotation angle relative to the axis rotation $\bm{v}=\frac{\bm{e}_z\times\bm{n}}{|\bm{e}_z\times\bm{n}|}$. This transformation satisfies}

\begin{align}
    U^\dagger (\bm{n} \cdot \bm{\sigma})U=\sigma_z.
\end{align}

\noindent {\color{black} Upon transforming the spinor as} $\psi\rightarrow\ \tilde{\psi}=U^\dagger\psi$, the derivatives acquire the SU(2) gauge field,

\begin{align}
    \partial_\mu \rightarrow D_\mu=\partial_\mu+\frac{iq}{\hbar} a_\mu,   \quad  a_\mu \equiv (i\hbar/q) U^\dagger\partial_\mu U
\end{align}

\noindent with $\mu=(x,y,z,t)$. {\color{black} The resulting equation of motion becomes} 

\begin{align}
      i\hbar \partial_t\tilde{\psi}= \left(\frac{(\bm{p} - q\bm{\mathcal{A}^e})^2}{2m} + q\mathcal{V}^e - J_{sd} \bm{n}(\bm{r},t) \cdot \bm{\sigma} \right)\tilde{\psi}
\end{align}

\noindent where $\bm{a}=(\bm{\mathcal{A}}^e,-\mathcal{V}^e)$ components are $2\times 2$ matrices in spin space. This shows that the electron couples minimally to the SU(2) gauge potential $a_\mu$.

 {\color{black} In adiabatic limit, the interband spin-flip processes are suppressed, and the gauge field can be projected onto the majority ($\uparrow$) or minority ($\downarrow$) spin band aligned with $\bm{n}$, yielding an Abelian U(1) Berry gauge potential $a^{(\sigma)}_\mu$,} 


\begin{align}
    a^{(\sigma)}_\mu &=(i\hbar/q)\langle \sigma|U^\dagger\partial_\mu U |\sigma\rangle \nonumber \\
    &=\pm (\hbar/e) (1-\cos \Theta)\partial_\mu\Phi
\end{align}

\noindent where $\sigma=\uparrow,\downarrow$ represents the majority and minority bands, respectively. Here, we have inserted the parameter $q=-e$ for the definition of an electron. Thus, a general gauge-invariant formula for the emergent magnetic and electric fields in terms of polar $\Theta$ and azimuthal $\Phi$ angles are defined as follow,
    
\begin{align}
\bm{B}^e(r,t) &= \bm{\nabla}\times \mathcal{\bm{A}}^e \nonumber\\
            &=s\frac{\hbar}{2e} \sin \Theta (\bm{\nabla}\Theta \times \bm{\nabla}\Phi), \label{eq:gauge-inv-B} \\
\bm{E}^e(r,t)&= -\partial_t \mathcal{\bm{A}}^e -\bm{\nabla}\mathcal{V}^e \nonumber \\
            &= s\frac{\hbar}{2e} \sin \Theta (\bm{\nabla}\Theta\partial_t\Phi-\bm{\nabla}\Phi\partial_t\Theta ).
\label{eq:gauge-inv-E}
\end{align}

\noindent The sign of $s=\pm$ stands for the majority (spin-up) and minority (spin-down) bands. The formula written in the polar and azimuthal angles is more convenient for our purpose in the present study. In the next section, we use the positive sign $s$ representing the majority band.

{\color{black} We emphasize here that Eqs. \ref{eq:gauge-inv-B} and \ref{eq:gauge-inv-E}  are valid only in the adiabatic strong-coupling regime, where the electron spin follows the slowly varying magnetization texture. Although the resulting expression for $E^e$ depends only on the spin texture and fundamental constants, this does not imply that the effect survives in the limit $J_{sd} \rightarrow 0$. In this limit, the adiabatic approximation breaks down and the emergent gauge-field description is no longer applicable.}

{\color{black} 
Although the emergent electric field arises locally from the geometric Berry phase associated with a smooth and time-dependent spin texture regardless of its topological character, a finite macroscopic response requires a nonvanishing skyrmion density. In a skyrmion lattice, the topological charge of each skyrmion ensures that the emergent fields add constructively upon coarse-graining. In contrast, non-topological textures or stripe-like phases generally yield vanishing averaged emergent fields due to cancellation of opposite contributions.
}

Note that the skyrmion profile is characterized by the radial dependence of the polar angle and the azimuthal dependence of the helicity. This profile is primarily determined by the equilibrium radius and the domain-wall width of skyrmion. In principle, the skyrmion radius~$R_0$ and wall width $\Delta_w$ can be directly extracted from experiments \cite{Meyer2019,Wu2021, Rohart_PhysRevB.88.184422, Wang2018}. Theoretically, these geometric parameters can be related to microscopic quantities \cite{Rohart_PhysRevB.88.184422, Wang2018}. For example, in the model of Ref.~\cite{Wang2018}, one finds

\begin{align}
    \Theta(r) &= 2 \arctan\left[\frac{\sinh(R_0/\Delta_w)}{\sinh(r/\Delta_w)}\right], \\
    R_0 &=\pi D\sqrt{\frac{A}{16AK^2-\pi^2D^2K}}, \\
    \Delta_w &=\frac{\pi D}{4K},
\end{align}

\noindent where $A$ is the exchange stiffness, $D$ the Dzyaloshinskii–Moriya interaction, and $K$ the anisotropy constant.

\section{Rigid Oscillation of Skyrmion} 
\subsection{Breathing Mode}

\begin{figure}[b]
    \centering
    \includegraphics[width=0.85\linewidth]{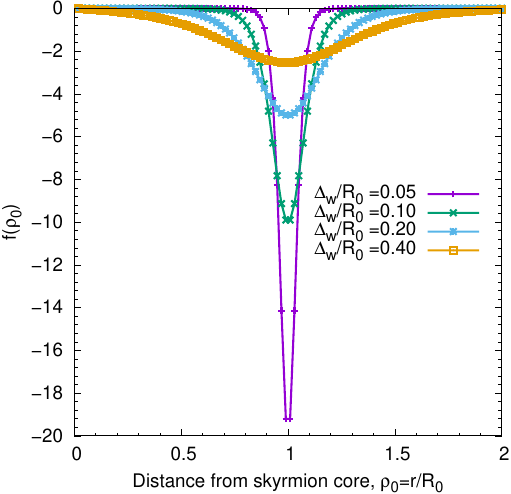}
    \caption{\color{black}Radial shape factor $f(\rho_0)=\sin\Theta(\rho_0) \Theta'(\rho_0)$ as a function of distance from the skyrmion core $\rho_0=r/R_0$ with skyrmion radius $R_0$ for different skyrmion wall width $\Delta_w$.}
    \label{fig:f0vsDw}
\end{figure}

Let us start from the ansatz {\color{black}for a single breathing skyrmion} with rigid oscillation, 

\begin{align}
\bm{n}(r,\varphi,t) =
\begin{pmatrix} \sin\Theta(\rho) \cos (v\varphi + \gamma) \\ \sin\Theta(\rho) \sin (v\varphi + \gamma) \\ \cos \Theta(\rho) \end{pmatrix}
\end{align}

\noindent with $\rho\equiv r/R(t)$ is the dimensionless radius, $R(t)$ is the time-dependent skyrmion radius (breathing mode). $\Theta(\rho)$ is the time-dependent polar angle with the conditions $\Theta(0)=\pi$ and $\Theta(\infty)=0$, while $\varphi$ is the standard azimuthal angle, and $\gamma$ is a constant helicity ($\gamma=0,\pi$ for Neel type, $\gamma=\pm \pi/2$ for Bloch type). And $v$ is the vorticity where $v=\pm 1$ stands for a standard skyrmion and anti-skyrmion. 

Solve the ansatz, with $\Theta=\Theta(\rho(t))$ and $\Phi=v\varphi+\gamma$, in polar coordinate, 
\begin{align}
    \partial_t \Theta &=- \frac{\rho \dot{R}}{R}\Theta'(\rho), \;\Theta'(\rho)<0 \nonumber \\
    \partial_t \Phi &= 0, \; \; \; \partial_\varphi\Phi=v 
\end{align}

\noindent Thus, the emergent electric field is obtained,

\begin{align}
    \bm{E}^e (r,t)&= \frac{v\hbar}{2e}  \frac{\dot{R}}{R^2}\sin \Theta(\rho) \Theta'(\rho) \bm{\hat{e}}_\varphi.
    \label{eq:breathing}
\end{align}

\noindent This shows that the emergent field in the breathing mode is purely circulating (azimuthal) and the amplitude is proportional to velocity $\dot{R}$ and scales by $1/R^2$. The radial shape {\color{black} of emergent field} is determined by $\sin \Theta(\rho) \Theta'(\rho)$, which is localized around the skyrmion wall where the polar angle $\Theta(\rho)$ changes rapidly. {\color{black} Fig. \ref{fig:f0vsDw} shows the radial shape factor $f(\rho_0)=\sin \Theta(\rho_0) \Theta'(\rho_0)$ as a function of distance from the skyrmion core $\rho_0=r/R_0$ with the skyrmion radius $R_0$ for different skyrmion wall width $\Delta_w$. The shape factor becomes more localized at narrow wall. } 

Now, we estimate the induced electromotive force induced by spin-texture dynamics or spin-motive force (smf) around a circle at radius $r$,

\begin{align}
    \varepsilon(r,t) &= \oint \bm{E}^e\cdot d\bm{l} \\
    &=\int_0^{2\pi} E_\varphi(r,t) rd\varphi=2\pi rE_\varphi(r,t) \\
    &=\frac{v\pi \hbar}{e} \frac{\dot{R}}{R}\rho \sin \Theta(\rho) \Theta'(\rho),
\end{align}

\noindent the smf is therefore localized around the skyrmion wall where $\sin \Theta(\rho) \Theta'(\rho)$ is considerably large.

Define the breathing oscillation, $R(t)=R_0+\delta R \cos(\Omega_B t)$, thus the amplitude of radial velocity $\dot{R}$,

\begin{align}
    \dot{R}_{\text{amp}}&=\Omega_B\delta R. 
\end{align}

\noindent By assuming the skyrmion wall width $\Delta_w$, {\color{black} the narrow-wall approximation yields}
\begin{align}
    \Theta'(\rho)=\frac{d\Theta(\rho)}{dr} \frac{dr}{d\rho}\approx -\frac{\pi R}{\Delta_w} \; \text{\color{black} },
\end{align}

\noindent thus, the smf amplitude is given by
\begin{align}
    \varepsilon_{\text{amp}}&\approx\frac{v\pi^2\hbar}{e} \frac{\Omega_B\delta R }{\Delta_w}\rho \sin \Theta(\rho). 
\end{align}

\noindent Suppose that {\color{black} the skyrmion ($v=+1$), which typically observed in metallic ferromagnet \cite{Fert2017_NatRevMat,Nagaosa2013_NatNano},} has the angular frequency $\Omega_B$ is $2\pi\times1$ GHz, skyrmion radius $R_0=~50$~nm, breathing amplitude $\delta R= 5$ nm ($\sim10\%R_0$), and skyrmion wall width $\Delta_w=10$ nm. Using the fundamental constant $\hbar/e=6.582 \times 10^{-16} \text{Vs}$, the spin-motive-force amplitude near the skyrmion wall ($\rho\approx1 \rightarrow\sin\Theta(1)\approx 1$) is estimated as $\varepsilon_{\text{amp}} \approx 20 \;\mu V$. {\color{black} The corresponding amplitude of the emergent electric field in the wall region is $|\bm{E}^e|=\frac{\hbar}{2e}\frac{\pi \Omega_B \delta R}{R_0 \Delta_w}\approx 6.5 \times 10^1 $~V/m.}

\subsection{\color{black}Rotational Modes}
 The ansatz for rotational CW/CCW modes of {\color{black}a single skyrmion} may be written as,
\begin{align}
\bm{n}(r,\varphi,t) =
\begin{pmatrix} \sin\Theta(\rho) \cos (v\varphi + \gamma + \Omega_G t) \\ \sin\Theta(\rho) \sin (v\varphi + \gamma + \Omega_G t) \\ \cos \Theta(\rho) \end{pmatrix},
\end{align}
\noindent here, $\rho\equiv r/R_0$ is the time-independent. We added the time-dependent rotation angle with angular velocity $\Omega_G$, in which $\Omega_G<0$ and $\Omega_G>0$ correspond to the clockwise (cw) and counterclockwise (ccw) rotations, respectively. 

\noindent For this ansatz, $\Theta=\Theta(\rho)$ and $\Phi=v\varphi+\gamma+\Omega_G t$,
\begin{align}
    \partial_r \Theta&=(1/R_0)\Theta'(\rho), \;\Theta'(\rho)<0, \nonumber\\
    \partial_t \Theta &= 0,  \;\;\; \partial_t \Phi = \Omega_G. 
\end{align}

\noindent The emergent electric field induced,

\begin{align}
    \bm{E}^e(r,t)=\frac{\hbar}{2e} \frac{\Omega_G}{R_0} \sin \Theta(\rho)\Theta'(\rho) \bm{\hat{e}}_r,
\end{align}

\noindent which is pure radial and non-oscillating field. {\color{black} Similar to breathing modes, the emergent field is proportional to $\sin \Theta(\rho) \Theta'(\rho)$, which is largest in the wall region.}

Next, we examine the radial potential difference between the skyrmion core and a point with radius $r$,

\begin{align}
    V(r,t)&=-\int_0^r E(r',t) dr'  \\
    &=-\frac{\hbar}{2e} \Omega_G \int_0^r \sin \Theta(\rho)\frac{d\Theta(\rho)}{dr'} dr' \\
    &=-\frac{\hbar}{2e}\Omega_G\int_{\Theta(0)}^{\Theta(r)} \sin \Theta d\Theta  \\
    &=-\frac{\hbar}{2e}\Omega_G (\cos\Theta(0)-\cos\Theta(r)),
\end{align}

\noindent if we measure between the skyrmion core ($\Theta(r=0)=\pi$) and far outside ($\Theta(r=\infty)=0$), the potential difference $V$ is,

\begin{align}
    V(r=0 \rightarrow \infty )=(\hbar/e) \Omega_G.
\end{align}

\noindent Thus, a rigid rotation of the skyrmion produces a radial voltage of magnitude $|\hbar\Omega_G/e|$ between core and far field; its sign flips with the rotation sense (sign of~$\Omega_G$). For instance, a small rotation with angular frequency of $\Omega=2\pi \times 1$ GHz yields a radial voltage with magnitude $|\hbar\Omega_G/e| \approx~4 \;\mu V$. {\color{black} While, the corresponding emergent electric field in the wall region with $\Delta_w=10\;\text{nm}$ is $|\bm{E}^e| = (\hbar/2e)(\pi |\Omega_G|/\Delta_w)\approx~6.5 \times 10^2 \;\text{V/m}.$} 

\subsection{Macroscopic field of symmetric skyrmion}
 Now consider the skyrmion lattice (SkX) in the dilute limit approximation, which can be constructed from the superposition of individual skyrmions~$\bm{n}$.  We assume that the overlapping tails among the skyrmions are negligible, i.e., the lattice spacing $a$ is much larger than the skyrmion diameter ($a \gg 2R_0$). {\color{black} Such a well-separated SkX could be realized in systems with strong perpendicular magnetic anisotropy \cite{Aji_2025}. In this regime, the SkX texture $\bm{m}(\bm{r},t)$ may be approximated by the ansatz}
 
\begin{align}
{\color{black}
    \bm{m}(\bm{r},t) } \approx\sum_j \bm{n}\left(\frac{\bm{r}-\bm{R_j}}{R(t)} \right). 
\end{align}

\noindent Each skyrmion with radial coordinate $\rho_j = \frac{\bm{r}-\bm{R_j}}{R(t)}$, contributes to the local emergent electric field with natural in-phase breathing oscillation,

\begin{align}
    \bm{E}^e_j (r,t)&= \frac{v\hbar}{2e}  \frac{\dot{R}}{R^2}\sin \Theta(\rho_j) \Theta'(\rho_j) \bm{\hat{e}}_\varphi.
    \label{eq:dilute_lattice}
\end{align}
\noindent {\color{black} These emergent fields are strongly localized in the skyrmion wall (domain-wall) region and decay rapidly both toward the skyrmion core and into the uniform background of the skyrmion lattice. } 

 {\color{black} We now define the coarse-grained (macroscopic) emergent electric field as the spatial average over a single skyrmion unit cell of area $A_{\rm{cell}}$,}
 
\begin{align}
\langle \bm{E}^e  \rangle &= \frac{1}{A_\mathrm{cell}}  \int_\mathrm{cell} {\color{black} \bm{E}^e [\bm{m}] } d^2r. 
\end{align}

\noindent {\color{black} For the symmetric and in-phase breathing mode, this average vanishes identically,}

{\color{black}
\begin{align}
        \langle \bm{E}^e \rangle = \frac{1}{A_\mathrm{cell}}  \int_0^{2\pi}\int_0^{r_{ws}} E^{e}_\varphi(r,t) \bm{\hat{e}}_\varphi \, rdr  
        d\varphi=0,
\end{align}

\noindent where $r_{ws}$ is the radial distance to the Wigner–Seitz boundary of the skyrmion lattice.}

{\color{black} This result indicates that the in-phase breathing of a symmetric skyrmion lattice generates a local, circulating emergent electric field, corresponding to a finite spin-motive force (EMF) around each skyrmion, while the unit-cell-averaged (macroscopic) electric field vanishes. Consequently, the breathing mode does not define a scalar two-probe voltage in an ideal symmetric lattice and is most naturally detected via loop or inductive measurement geometries that probe the EMF.

In a similar way, owing to parity symmetry, the emergent electric field generated by rotational modes, although predominantly radial and capable of producing a well-defined local voltage between the skyrmion core and the far field, also yields a vanishing unit-cell-averaged electric field in a perfectly symmetric lattice. Thus, in both breathing and rotational modes, the macroscopic emergent field obtained by coarse-graining is zero in the absence of additional symmetry breaking.

A finite unit-cell-averaged emergent electric field generally may arise only when spatial symmetries are broken, for example by lattice anisotropy, strain, boundaries, or coupling to phonons, or when phase differences develop between the polar and azimuthal components of the texture dynamics. In such cases, the resulting coarse-grained emergent field acts as an effective driving field in electronic transport, giving rise to a longitudinal charge current density $j_\text{c}$. For instance, in a two-current description, majority ($\uparrow$) and minority ($\downarrow$) conduction electrons experience effective fields $+\bm{E}^e_{eff}$ and $-\bm{E}^e_{eff}$, respectively, giving rise to spin-resolved currents $\bm{j}_\uparrow=\sigma_\uparrow \bm{E}_{eff}^e$ and $\bm{j}_\downarrow=-\sigma_\downarrow \bm{E}_{eff}^e$. The net charge current density is obtained by summing these contributions,

\begin{align}
    \bm{j}_\text{c}=\bm{j}_\uparrow+\bm{j}_\downarrow=\sigma_{eff} \bm{E}_{eff}^e,
\end{align}

\noindent where $\sigma_{eff}=\sigma_\uparrow - \sigma_\downarrow$ is the effective conductivity of spin-polarized conduction electrons. This relation has the standard Ohmic form and describes the longitudinal charge current driven by the coarse-grained emergent electric field. In the following section, we discuss how lattice deformation and phonon coupling provide concrete mechanisms for generating a finite coarse-grained emergent electric field.}



\section{Magnetoelastic coupling}
To capture the effect of lattice deformation on the induced macroscopic electric field, a small lattice perturbation via magnetoelastic coupling needs to be introduced to the static/equilibrium skyrmion texture. The total energy coming from the static skyrmion and additional magnetoelastic coupling may be written as \cite{Lu_PhysRevB.91.100405,Callen_PhysRev.129.578},
\begin{align}
    E[\bm{m},\bm{\varepsilon}]=\int \mathcal{H}_{mag}[\bm{m}]+\mathcal{H}_{me}[\bm{m},\bm{\varepsilon}] \;d^2r,
\end{align}

\noindent {\color{black} where $\mathcal{H}_{mag}[\bm{m}]$ denotes the magnetic energy density of equilibrium skyrmion texture, } which typically originates from the exchange, DMI, anisotropy, and Zeeman terms. While the magnetoelastic energy $\mathcal{H}_{me}[\bm{m},\bm{\varepsilon}]$ is given by,
\begin{align}
\mathcal{H}_{me}[\bm{m},\bm{\varepsilon}]=B_{ijkl}\varepsilon_{ij}m_km_l,
\end{align}
\noindent with the normalized magnetization $\bm{m}$, strain tensor $\varepsilon_{ij}$, and magnetoelastic coupling constants $B_{ijkl}$.

 Define the skyrmion radius with a small boundary distortions with mode amplitude $a_\ell$,
\begin{align}
    R(\varphi,t)=R_0+\sum_\ell a_\ell(t)\psi_\ell(\varphi),
\end{align}

{\color{black}
\noindent with $\psi(\varphi)=\cos(\ell(\varphi-\varphi_\ell))$ is angular dependence of a particular mode $\ell$ of deformation. The fundamental modes correspond to isotropic ($\ell=0$), dipolar ($\ell=1$), elliptical ($\ell=2$), triangular ($\ell=3$), etc. Each harmonic $\ell$ has $C_\ell$ symmetry in-plane (invariant under rotation $\varphi\rightarrow\varphi+2\pi/\ell$), so it is naturally tied to lattices/environments whose symmetry allows $C_\ell$. 
Expanding the total energy functional around the equilibrium to linear order in the mode amplitudes $a_\ell$ yields
}
 
\begin{align}
H_\text{int}&=-\sum_\ell a_\ell F_\ell, \\
F_\ell &=- \frac{\partial E}{\partial a_\ell}|_{a_\ell=0}.
\end{align}

 A standard equation of motion of damped harmonic dynamics for the displacement $a_\ell$ derived from Euler-Lagrange formalism by including a Berry phase kinetic term, magnetic potential, and phenomenological damping can be written as follows \cite{Keever_PhysRevB.99.054430,Lin_PhysRevB.96.014407,Büttner2015},

\begin{align}
    \mu_\ell \ddot{a}_\ell + \gamma_\ell\dot{a}_\ell+\kappa_\ell a_\ell&=F_\ell(\bm{r},t),
    \label{eq:equationofmotion}
\end{align}
\noindent where the parameters $\mu_\ell$, $\gamma_\ell$, $\kappa_\ell$ are the effective inertia, damping, and stiffness, respectively. While the driving force $F_\ell(\bm{r},t)$ owing to the magnetoelastic energy,
\begin{align}
    F_\ell(\bm{r},t) &= \int \Lambda_{ij}^{(\ell)}(\bm{r'};R_0)\varepsilon_{ij}(\bm{r+r'},t) \; d^2r',\\
    {\color{black}
    \Lambda_{ij}^{(\ell)}(\bm{r'};R_0)} &= {\color{black}-B_{ijkl}\frac{\partial}{\partial{a_\ell}}(m_km_l)|_{a_\ell=0}, } 
\end{align}

\noindent with a single coherent phonon in the strain field $\varepsilon_{ij}$, 
\begin{align}
    \varepsilon_{ij}(\bm{r+r'},t)=\varepsilon_{ij}^{(0)} \cos(\bm{q\cdot} (\bm{r}+\bm{r'}) -\omega t).
\end{align}

 \noindent At a small fluctuation around the skyrmion center $\bm{r}$, the magnetoelastic force may be approximated as, 

\begin{align}
    F_\ell(\bm{r},t) &\approx \mathcal{G}_{ij}^{(\ell)}(R_0)\varepsilon_{ij}^0\cos(\bm{q}\cdot \bm{r}-\omega t), \\
    \mathcal{G}_{ij}^{(\ell)}(R_0)&=\int \Lambda_{ij}^{(\ell)}(\bm{r'};R_0)  d^2r'.
\end{align}

{\color{black}
 Typical metallic ferromagnets used in skyrmion multilayers have magnetoelastic constants of order $B_{eff}^{(3D)} \sim 10^6-10^7$ J/m$^3$ (e.g., Ni and CoFeB) \cite{Li2020}. For a film thickness $t_d\simeq 10 \;\text{nm}$, this corresponds to an effective areal constant $B_{eff}^{(2D)}=B_{eff}^{(3D)} t_d \simeq  10^{-2}-10^{-1} \;\text{J/m}^2$. Since the mode-induced magnetization variation is primarily concentrated in the skyrmion-wall region of width~$\Delta_w$, the coupling kernel can be approximated as $\Lambda_{ij}^{(\ell)}\sim B_{eff}^{(2D)}/\Delta_w$ with integrated area $2\pi R_0\Delta_w$, thus the effective coupling $\mathcal{G}_{ij}^{(\ell)}$ scales as
\begin{align}
   \mathcal{G}_{ij}^{(\ell)}\sim 2\pi R_0 B_{eff}^{(2D)} \beta_{ij}^{(\ell)},
\end{align}

\noindent yielding $\mathcal{G}^{(\ell)}_{ij}\sim 10^{-11}-10^{-8}$ N for $R_0\simeq 50$ nm and taking a representative range of $\beta_{ij}^{(\ell)} \sim 10^{-2}-1$ as dimensionless overlap factor determined by which strain component $(ij)$ matches the mode symmetry~$\ell$.
}

 The steady solution in the equation of motion of the standard driven-oscillator in Eq. \ref{eq:equationofmotion} gives

\begin{align}
    a_\ell(\bm{r},t)=|\tilde\chi_\ell(\omega)|\mathcal{G}_{ij}^{(\ell)}(R_0)\varepsilon_{ij}^0\cos(\bm{q}\cdot \bm{r}-\omega t+\phi_\ell),
\end{align}

\noindent with the Lorentzian susceptibility,
\begin{align}
    \tilde\chi_\ell(\omega)&=\frac{1}{\kappa_\ell-\mu_\ell\omega^2-i\gamma_\ell \omega}=|\tilde\chi_\ell|e^{i\phi_\ell}, \\
    |\tilde\chi_\ell|&=\frac{1}{\sqrt{(\kappa_\ell-\mu_\ell\omega^2)^2+(\gamma_\ell\omega)^2}}, \\
    \phi_\ell &=\tan^{-1}\left(\frac{-\gamma_\ell\omega}{\kappa_\ell-\mu_\ell\omega^2}\right),
\end{align}

\noindent here, the amplitude $|\tilde\chi_\ell|$ is huge near the mode's resonance, $\omega\simeq\sqrt{\kappa_\ell/\mu_\ell}$.

\begin{figure*}[t]
    \centering
    \includegraphics[width=\linewidth]{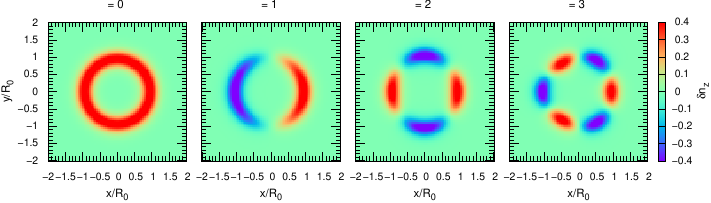}
    \caption{\color{black}Spatial distribution of the magnetization variation in the breathing mode, $\delta n_z=n_z(t)-n_z(0)$, evaluated at $t=\tau/4$, where $\tau$ is the oscillation period, for representative deformations: (a) isotropic, (b) dipolar, (c) elliptical, and (d) triangular.}
    \label{fig:dnz_maps}
\end{figure*}

 Finally, we derive the skyrmion radius under a small lattice perturbation due to a coherent-acoustic phonon coupled with the lattice deformation,

\begin{align}
    R({\varphi,\bm{r},t})&=R_0\left[1+\sum_\ell {\color{black} \tilde{\eta}_\ell } \cos(\bm{q}\cdot \bm{r}-\omega t+\phi_\ell) \right], \\
    {\color{black} \tilde{\eta}_\ell }&=\frac{|\tilde\chi_\ell| \mathcal{G}_{ij}^{(\ell)}\varepsilon_{ij}^0}{R_0}\psi_\ell(\varphi).
\end{align}

{\color{black} \noindent The index $\ell$ labels the azimuthal harmonic of the boundary deformation, $\tilde{\eta}_\ell$, and therefore determines which angular component of the strain field can drive the corresponding mode through the linear magnetoelastic coupling. Decomposing the in-plane strain into isotropic $\varepsilon_\text{iso}=(\varepsilon_{xx}+\varepsilon_{yy})/2$, deviatoric $\varepsilon_\text{dev}=(\varepsilon_{xx}-\varepsilon_{yy})/2$, and shear components $\varepsilon_\text{sh}=\varepsilon_{xy}$ \cite{LandauLifshitz_TheoryElasticity,Kuhl_ContinuumMechanicsNotes}, one finds that the uniform breathing mode ($\ell=0$) couples most directly to dilatational strain $\varepsilon_\text{iso}$, while the elliptical mode ($\ell=2$) is driven by anisotropic and shear strains $\varepsilon_\text{dev}$ and $\varepsilon_\text{sh}$. For an ideal circular skyrmion subject to a spatially uniform plane-wave strain, odd-$\ell$ modes (e.g., $\ell=1,3$) are symmetry-suppressed; they can be activated by finite-$q$ strain gradients, boundaries/defects, or patterned acoustic driving that breaks the local inversion symmetry around a skyrmion center \cite{Zhang2018,Yang2024}. For an order-of-magnitude estimate of $\tilde{\eta}_\ell$, we assume that the strain amplitude induced by coherent-acoustic wave on the order of $\varepsilon^{(0)}_{ij}\sim 10^{-4}$ \cite{Foerster2017,VANCAPEL201536}, a mode susceptibility $|\tilde\chi_\ell|\sim 10^3 $ m/N, an effective coupling $\mathcal{G}\sim 10^{-8}$~N, and a skyrmion radius $R_0=50$~nm. These values give a boundary-deformation amplitude $\tilde{\eta}_\ell\sim 0.02 \ll 1$.}

At finite temperature or in the presence of magnon–magnon interactions, the lattice deformation may involve superpositions or nonlinear mixing of multiple harmonic components \cite{Zheng_10.1063/5.0152543,Liu_PhysRevB.110.184413,Lan2025}. This harmonic mixing components have also an essential role in generating finite coarse-grained (dc or low frequency) electric field in the second-order corrections, as describe in the next section. 


\section{Coarse-Grained Electric Field}

\begin{figure*}
    \centering
    \includegraphics[width=0.95\linewidth]{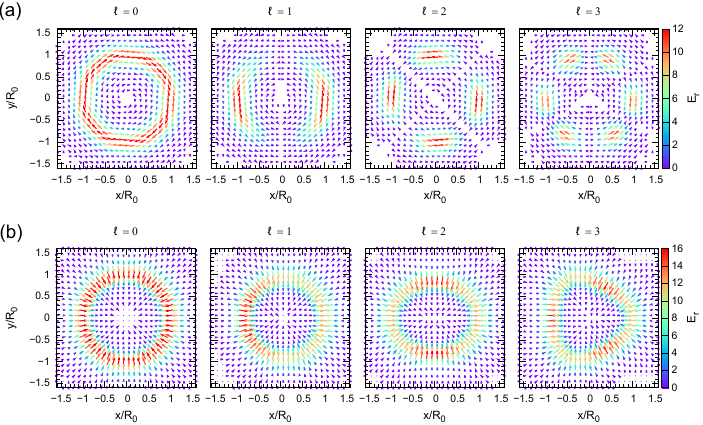}
    \caption{\color{black}Snapshots of normalized emergent electric field distribution with isotropic ($\ell=0$), dipolar ($\ell=1$), elliptical ($\ell=2$), and triangular ($\ell=3$) deformations at the first leading order $\mathcal{O}(\eta)$ for (a) breathing and (b) rotational modes. The intensity represents the strength of the electric field, $E_r = \sqrt{E'^2_x+E'^2_y}$.}
    \label{fig:field_maps}
\end{figure*}

\subsection{Breathing Mode}
\bo{\color{black} Angular-harmonic deformation}. Here, we first examine the case for pure breathing oscillation without spatial coupling through an acoustic phonon. The general lattice distortions in the breathing skyrmion which contain isotropic ($\ell=0$), dipolar ($\ell=1$), elliptical ($\ell=2$), triangular ($\ell=3$), and other higher harmonic terms may be written as 

\begin{align}
    R(\varphi,t)=R_0\left[1+ \mathcal{D}(\varphi,t)\right].
\end{align}

\noindent where $\mathcal{D}$ is the sum of the isotropic breathing oscillation $\ell=0$ and dynamical lattice distortions $\ell\geq1$,

\begin{align}
    \mathcal{D}(\varphi,t)=\sum_{\ell\geq0}\eta_\ell(t) \cos (\ell(\varphi-\varphi_\ell)), 
\end{align}

\noindent here, $\eta_\ell(t)=\eta_{0\ell}\cos(\omega_\ell t + \phi_\ell)$ is a time-dependent lattice distortion parameter in $\ell$-th harmonic order, $|\eta_{0\ell}|\ll 1$, having breathing frequency $\omega_\ell$ and phase factor $\phi_\ell$. {\color{black}The spatial distribution of magnetization variation, $\delta n_z~=~n_z(t)-n_z(0)$, after a quarter period of time $t=\tau/4$ for isotropic, dipolar, elliptical, and triangular deformations in breathing mode is shown in Fig.~\ref{fig:dnz_maps}.}



 Expanding the electric field in Eq. \ref{eq:breathing} for a small lattice distortion up to the first leading order of $\mathcal{D}$ gives

\begin{align}
    \bm{E}^e(r,\varphi,t)&=\frac{v\hbar}{2e} \frac{\dot{R}(\varphi,t)}{R_0^2}\left[f(\rho_0) - \mathcal{F}(\rho_0)\mathcal{D}(\varphi,t) \right]   \bm{\hat{e}}_\varphi,
    \label{eq:E_latdis}
\end{align}

\noindent with $f(\rho_0)= \sin \Theta(\rho_0) \Theta'(\rho_0)$, dimensionless parameter $\rho_0=r/R_0$, and the shape factor $\mathcal{F}(\rho_0)$,

    \begin{align}
        \mathcal{F}(\rho_0) = 2f(\rho_0)+\rho_0f'(\rho_0). 
    \end{align}

\noindent The resulting electric field is still purely circulating field similar to the rigid one, but here $\bm{E}^e$ carries an angular dependence of deformation $\varphi$. {\color{black} The electric field distribution at the first leading order $\mathcal{O}(\eta)$ for representative deformations are shown in Fig. \ref{fig:field_maps}(a).}

Now, we examine the induced macroscopic electric field owing to this angular dependence of deformation. 
It is convenient to convert the equation into the Cartesian coordinate $\hat{e}_\varphi=(-\sin\varphi,\cos\varphi)$. The spatial average of the first-order term in Eq. \ref{eq:E_latdis} over azimuthal angle $\varphi$, $\frac{1}{2\pi}\int_0^{2\pi} \cos(\ell(\varphi-\varphi_\ell))(-\sin\varphi,\cos\varphi) d\varphi$ gives non-zero only if $\ell=1$ (dipolar bias) otherwise the average vanishes. Thus, the nonzero macroscopic electric field over unit cell area contributed only from the dipolar bias in the first-order term $\mathcal{O}(\eta)$ is written as follows, 

\begin{align}
    \langle \bm{E}_1^e\rangle(t)=\mathcal{C}_f\dot{\eta}_1(t) \bm{\hat{e}}_{\varphi_1}
    \label{eq:breathing_dipolar}
\end{align}

\noindent with the constant $\mathcal{C}_f$,
\begin{align}
    \mathcal{C}_f=\frac{v\hbar}{2e}{\color{black} \frac{\pi R_0}{ A_\text{cell}} }\int_0^{\color{black} \rho_{ws} }  f(\rho_0) \rho_0d\rho_0,
\end{align}

{\color{black}
\noindent where $\rho_{ws}=r_{ws}/R_0>1$ denotes the (dimensionless) effective Wigner–Seitz radius of the unit-cell in a periodic skyrmion lattice. This result shows that the coarse-grained oscillating field appears along the dipolar axis $\bm{\hat{e}}_{\varphi_1}$, with the amplitude is given by $|\eta_{01}\omega_1\mathcal{C}_f|$. In the dilute limit $a\gg2R_0$, the emergent-field weight $f(\rho_0)$ is strongly localized near the skyrmion wall $\rho_0\simeq 1$ with characteristic width $\sim \Delta_w/R_0$. Therefore, the radial integral insensitive to the detailed unit-cell boundary and the lattice geometry enters primarily through the prefactor $1/A_\text{cell}$ (e.g., 
$A_\text{cell}=(\sqrt{3}/2)a^2$ for a triangular SkX). 

In the narrow-wall limit, where $f(\rho_0)$ is sharply localized near $\rho_0\simeq1$, the radial integral saturates to 
\begin{align}
    \int_0^{\rho_{ws}} f(\rho_0) \rho_0d\rho_0 \simeq -2.
\end{align}
 
 \noindent For a triangular SkX with lattice constant $a\approx4R_0$, with $R_0 = 50$~nm, the unit cell area is $A_\text{cell}\approx3.46 \times 10^{-14} \text{m}^2$. Taking a dipolar deformation amplitude $\eta_{01}=0.02$, and a drive frequency 1 GHz, the coarse-grained emergent-field amplitude in Eq. \ref{eq:breathing_dipolar} yields  $\sim 0.4$ V/m. For a measurement path $L\gg a$, where microscopic variations average out and the coarse-grained field may be treated as approximately uniform, the voltage along the dipolar axis is estimated as $V_\text{amp}\simeq |\langle E_1^e \rangle| L\sim 4 \; \mu$V for $L=10 \;\mu$m.} However, note that the time average of this field vanishes since it contains $\sin(\omega_1t+\phi_1)$, thus the macroscopic DC field vanishes, 
\begin{align}
\overline{\langle \bm{E}^e_1\rangle}^t=0    
\end{align}

 On the other hand, the second-order contribution $\mathcal{O}(\eta^2)$ contains $\dot{R}\mathcal{D}\equiv R_0\sum_{\ell,\ell'}\dot{\eta}_\ell\eta_{\ell'} \cos(\ell(\varphi-\varphi_\ell))\cos(\ell'(\varphi-\varphi_{\ell'}))$. For simplicity, assume that the phase factors are equal, $\varphi_\ell=\varphi_{\ell'}=\varphi_d$. The average of the product of $\dot{R}\mathcal{D}$ over azimuthal angle $\varphi$ gives non-zero values under some conditions. Let $\bm{M}(\varphi)=\cos(\ell(\varphi-\varphi_d))\cos({\ell'}(\varphi-\varphi_d))\bm{\hat{e}}_\varphi$, the average over $\varphi$ yields,

 \begin{align}
     \langle \bm{M}(\varphi) \rangle_\varphi = \frac{(\delta_{|\ell-\ell'|,1}+\delta_{\ell+\ell',1})}{4}\bm{\hat{e}}_{\varphi_d}.
 \end{align}

\noindent  Thus, the cell average of the second-order electric-field along the in-plane anisotropy direction $\bm{\hat{e}}_{\varphi_d}$ gives

\begin{align}
    \langle \bm{E}^e_{2} \rangle (t)&= -\frac{\mathcal{C}_\mathcal{F}}{4}\sum_{\ell,\ell'}\dot{\eta}_\ell\eta_{\ell'}(\delta_{|\ell-\ell'|,1}+\delta_{\ell+\ell',1})\bm{\hat{e}}_{\varphi_d}, \\
    \mathcal{C}_\mathcal{F}&=\frac{v\hbar}{e}{\color{black} \frac{\pi R_0}{A_\text{cell}} }\int_0^{ \color{black} \rho_{ws} } \mathcal{F}(\rho_0) \rho_0d\rho_0. 
\end{align}

\noindent Please note that $\int_0^{\rho_{ws}} \mathcal{F}(\rho_0)\rho_0d\rho_0$ in a constant $\mathcal{C}_\mathcal{F}$ vanishes for an infinite plane/periodic cell with no edge term, resulting in a vanishing second-order term.

{\color{black} In a finite system, however, the coarse-graining involves an effective Wigner–Seitz cutoff $\rho_{ws}=\Lambda$ (cell radius = $\Lambda R_0$), thus the radial integral remains finite, 

\begin{align}
    \int_0^{\rho_{ws}} \mathcal{F}(\rho_0)\rho_0d\rho_0=[\rho_0^2f(\rho_0)]_0^\Lambda=\Lambda^2 f(\Lambda).
\end{align}

\noindent In the case of narrow wall with cutoff radius $\Lambda\rightarrow1$ (near wall), the spatial profile is $f(1)\approx-\frac{\pi R_0}{\Delta_w}$. Using circular cell approximation, $A_\text{cell}\approx\pi (\Lambda R_0)^2$, the second-order coarse-grained emergent field reduces to

\begin{align}
  \langle E_2^e\rangle(t) \simeq \frac{v\hbar}{4e}\frac{\pi}{\Delta_w}\sum_{\ell,\ell'}\dot{\eta}_\ell\eta_{\ell'}(\delta_{|\ell-\ell'|,1}+\delta_{\ell+\ell',1})\bm{\hat{e}}_{\varphi_d}.  
\end{align}
 } %
 
\noindent This result suggests that the nonzero contribution in the second-order macroscopic-field, in finite systems with edge term, originates either from breathing-dipolar interaction ($\ell=1, \ell'=0$ or $\ell=0,\ell'=1$) or from the interaction between two adjacent higher harmonic terms ($|\ell-\ell'|=1$). For instance, the coupling between dipolar bias with elliptical distortion could also result in nonzero second-order coarse-grained oscillating electric field. The elliptical distortion cannot solely induce such a coarse-grained field. {\color{black} As an order-of-magnitude estimate, we consider two coherent deformations (e.g., dipolar and elliptical) with equal amplitudes $\eta_{01}=\eta_{02}=0.02$ driven at frequency 1 GHz. For $R_0=50$ nm and $\Delta_w=10$ nm, the second-order coarse-grained emergent-field amplitude is estimated as $|\langle E_2^e\rangle| \simeq 6.5 \times 10^{-2}$ V/m. }

Furthermore, the second-order term is also proportional to $\dot{\eta}_\ell \eta_{\ell'}\equiv-\omega_\ell\eta_{0\ell}\eta_{0\ell'} \sin(\omega_\ell t+\phi_\ell)\cos(\omega_{\ell'}t+\phi_{\ell'})$. The time average of the sinusoidal product does not vanish in the case $\omega_\ell=\omega_{\ell'}=\Omega_B$, yields $\overline{\dot{\eta}_\ell \eta_{\ell'}}^t=-\frac{1}{2}\Omega_B\eta_{0\ell}\eta_{0\ell'}\sin(\phi_\ell-\phi_{\ell'})$. Thus, the microscopic DC electric field induced from the second-order correction is derived as,

\begin{align}
    \overline{\langle \bm{E}^e_{2} \rangle}^t &= \mathcal{C}'_\mathcal{F} \sum_{\ell,{\ell'}}\sin(\phi_\ell-\phi_{\ell'})(\delta_{|\ell-\ell'|,1}+\delta_{\ell+\ell',1})\bm{\hat{e}}_{\varphi_d}, \\
    \mathcal{C}'_\mathcal{F}&={\color{black} \frac{1}{8} }\Omega_B\eta_{0\ell}\eta_{0\ell'}\mathcal{C}_\mathcal{F}.
\end{align}

\noindent It implies that a finite DC macroscopic field requires phase lags between two harmonic mixing (e.g., breathing-dipole with $\phi_1-\phi_0\neq0$), or more generally between adjacent $\ell$ and $\ell+1$ components. Symmetry alone (without phase lag) gives zero DC fields.

\bo{Phonon coupling}. Now, we consider the spatial {\color{black}modulation} owing to the acoustic phonon with momentum $\bm{q}$, frequency $\Omega_p$, {\color{black}and phase factor $\chi_\ell$}. The skyrmion deformation $\mathcal{D}$ in this case is modified as follows,

\begin{align}
    \mathcal{D}(\bm{r},\varphi,t)=\sum_{\ell\geq0} \eta_\ell(t) \cos(\ell(\varphi-\varphi_\ell))\cos(\bm{q}\cdot\bm{r}-\Omega_p t+\chi_\ell)
\end{align}

\noindent The induced electric field up to the second order of $\eta$,

\begin{align}
    \bm{E}^e(\bm{r},\varphi,t)=\frac{v\hbar}{2e}\frac{1}{R_0}[f(\rho_0)\mathcal{\dot{D}}-\mathcal{F}(\rho_0)\mathcal{D}\mathcal{\dot{D}}] \bm{\hat{e}}_\varphi,
\end{align}

\noindent define $\theta=\varphi-\varphi_q$ and use the following vector projection,

\begin{align}
    \bm{\hat{e}}_\varphi(\varphi)=\cos{(\theta)}\bm{\hat{e}}_\varphi(\varphi_q)+\sin{(\theta)}\bm{\hat{e}}_\varphi(\varphi_q+\frac{\pi}{2}).
\end{align}

\noindent Let us first examine the first-order macroscopic field, 

\begin{align}
    \langle \bm{E}_1^e\rangle=\mathcal{C}_f[C(r,t)\bm{\hat{e}}_{\varphi_q}+S(r,t)\bm{\hat{e}}_{\varphi_q+\pi/2}],
    \label{eq:breath_1}
\end{align}

\noindent with coefficients of $C$ and $S$,
\begin{align}
    C(r,t)&=\frac{1}{\pi}\int_0^{2\pi} \cos(\varphi-\varphi_q)g(\varphi,r,t)d\varphi,\\
    S(r,t)&=\frac{1}{\pi}\int_0^{2\pi} \sin(\varphi-\varphi_q)g(\varphi,r,t)d\varphi,
\end{align}

\noindent and $g$ factor is defined, 
\begin{align}
    g&\equiv \dot{\mathcal{D}}=\cos(\ell(\varphi-\varphi_\ell))[\dot{\eta}_\ell\cos(a\cos(\theta)+b_\ell) \nonumber \\
    &\quad +\Omega_p\eta_\ell\sin(a\cos(\theta)+b_\ell)],
\end{align}
\noindent the following parameters have been also defined, $a_q= qr$ and $b_\ell=-\Omega_pt+\chi_\ell$. We consider in the case of long-wavelength limit ($a_q=qr\ll1$), the approximation gives
\begin{align}
    \cos(a_q\cos(\theta)+b_\ell)&=\cos(b_\ell) - a_q\cos(\theta)\sin(b_\ell)+\mathcal{O}(a_q^2), \\
    \sin(a_q\cos(\theta)+b_\ell)&=\sin(b_\ell) + a_q\cos(\theta)\cos(b_\ell)+\mathcal{O}(a_q^2).
\end{align}

\noindent Solve the azimuthal integral part at long-wavelength limit and apply the following useful vector projector,
\begin{align}
    \cos(\Delta)\bm{\hat{e}}_{\varphi_q} + \sin(\Delta) \bm{\hat{e}}_{\varphi_q+\frac{\pi}{2}} = \bm{\hat{e}}_{\varphi_q+\Delta}.
\end{align}

\noindent The nonzero contribution of electric field at long-wavelength limit up to the first leading order of $q$ yields,

\begin{align}
    \langle \bm{E}_1^e\rangle&=\mathcal{C}^{(0)}_f[\dot{\eta}_1\cos(b_1)+\Omega_p\eta_1\sin(b_1)]\bm{\hat{e}}_{\varphi_1} \nonumber\\
    & \quad -\mathcal{C}^{(q)}_f[\dot{\eta}_0 \sin(b_0)-\Omega_p\eta_0 \cos(b_0)] \bm{\hat{e}}_{\varphi_q} \nonumber \\
    & \quad -\frac{\mathcal{C}^{(q)}_f}{2}[\dot{\eta}_2 \sin(b_2)-\Omega_p\eta_2 \cos(b_2)] \bm{\hat{e}}_{2\varphi_2-\varphi_q},
\end{align} 

\noindent with the constants $\mathcal{C}^{(0)}_f$ and $\mathcal{C}^{(q)}_f$ 
\begin{align}
    \mathcal{C}^{(0)}_f &=\frac{v\hbar}{2e} {\color{black} \frac{\pi R_0} {A_\text{cell}} }\int_0^{\color{black}\rho_{ws}} f(\rho_0)\rho_0d\rho_0, \nonumber \\
    \mathcal{C}^{(q)}_f &= \frac{v\hbar}{2e}{\color{black} \frac{\pi q R_0^2} {A_\text{cell}} } \int_0^{\color{black}\rho_{ws}} f(\rho_0)\rho_0^2d\rho_0,
    \label{eq:Cfq}
\end{align}

\noindent which do not vanish, at infinite systems, for any reasonable skyrmion wall profiles. 

{\color{black} This result indicates that the leading macroscopic oscillating field at $\mathcal{O}(\eta)$ originates from the dipolar bias $\eta_1$ even in the long-wavelength limit ($q\rightarrow0$). By contrast, the isotropic breathing 
$\eta_0$ and elliptical distortion $\eta_2$ enter only through the first-order gradient correction $\mathcal{O}(q)$ via the phonon-induced lattice modulation.}

While, if we introduce a harmonic time dependence $\dot{\eta}_\ell=-\Omega_B\eta_{0\ell}\sin(\Omega_B t+\phi_\ell)$, we find that the dc component of the macroscopic emergent electric field vanishes, even when the breathing harmonics (isotropic/dipolar/elliptical) and the coherent acoustic phonon carry different phases. This conclusion is independent of the relative frequencies and extends to higher-order terms in $q$,


\begin{align}
    \overline{\langle \bm{E}^e_1 \rangle}^t = 0.
\end{align}

 On the other hand, the second-order correction $\mathcal{O}(\eta^2)$ can be reconstructed, in similar way, from Eq. \ref{eq:breath_1}, by replacing $g\rightarrow g'\equiv \mathcal{D}\dot{\mathcal{D}}$ and $\mathcal{C}_f\rightarrow -\mathcal{C}_\mathcal{F}$. {{\color{black} For simplicity we assume phase-locked phonon coupling $b_\ell=b_{\ell'}=b_p$}, the second-order correction ($\mathcal{O}(\eta^2)$) in the electric field at long-wavelength limit and linear in $q$ is derived as follows,

\begin{align}
    \langle \bm{E}_2^e\rangle &=\frac{\pi\mathcal{C}_\mathcal{F}^{(q)}}{4}\sum_{\ell,\ell'} \{[\eta_\ell\dot{\eta}_{\ell'}\sin(2b_p)-\Omega_p\eta_\ell\eta_{\ell'}\cos(2b_p)] \nonumber \\
    &\quad \times [2\delta_{{\ell},{\ell'}}\cos(\psi_-) \bm{\hat{e}}_{\varphi_q}+\delta_{|{\ell}-{\ell'}|,2}\bm{\hat{e}}_{\varphi_q+\text{sgn}({\ell}-{\ell'})\psi_-} \nonumber\\
    & \quad+\delta_{\ell+\ell',2}\bm{\hat{e}}_{\varphi_q+\psi_+}]
    \}
\end{align}

\noindent with phase $\psi_\pm=\ell\Delta_
\ell\pm \ell'\Delta_{\ell'}$, $\Delta_\ell=\varphi_\ell-\varphi_q$, and modified shape factor $\mathcal{C}^{(q)}_{\mathcal{F}}$,

\begin{align}
    \mathcal{C}^{(q)}_{\mathcal{F}}=\frac{v\hbar}{2e}{\color{black} \frac{\pi qR_0^2}{A_\text{cell}} }\int_0^{\color{black} \rho_{ws}} \mathcal{F}(\rho_0)\rho_0^2d\rho_0.
    \label{eq:CFq}
\end{align}

\noindent This implies that the second-order correction of oscillating macroscopic field comes from the self mixing $\delta_{\ell,\ell'} $, second-nearest mixing $\delta_{|
\ell-\ell'|,2}$, and other contributions $\delta_{\ell+\ell',2}$. 

At quasi-static phonon drive where the oscillation period of phonon is much longer than that of spin dynamics, $\Omega_p\ll\Omega_B$. {\color{black}In such case, $b_p$ is effectively constant}. Let $\eta(t)=\eta_{0\ell}\cos(\Omega_Bt+\phi_\ell)$, then take the time average from the spin drive ({\color{black}$t_B\sim2\pi/\Omega_B$}), $\overline{\eta_\ell\dot{\eta}_{\ell'}}^{t_B}=\frac{\Omega_B}{2}\eta_{0\ell}\eta_{0{\ell'}}\sin(\phi_\ell-\phi_{\ell'})$, and $\overline{\eta_\ell\eta_{\ell'}}^{t_B}=\frac{1}{2}\eta_{0\ell}\eta_{0{\ell'}}\cos(\phi_\ell-\phi_{\ell'})$. The nonzero macroscopic field for the fast spin drive or quasi-static phonon in the $\mathcal{O}(q)$ order gives

\begin{align}
    \overline{\langle \bm{E}_2^e\rangle}^{t_B} &=\frac{\pi\mathcal{C}_\mathcal{F}^{(q)}}{4}\sum_{\ell,{\ell'}} \{[\frac{\Omega_B}{2}\eta_{0\ell}\eta_{0{\ell'}}\sin(\phi_\ell-\phi_{\ell'})\sin(2b_p) \nonumber\\ 
    &\quad -\frac{\Omega_p}{2}\eta_{0\ell}\eta_{0{\ell'}}\cos(\phi_\ell-\phi_{\ell'})\cos(2b_p)] \nonumber \\
    &\quad \times [2\delta_{\ell,{\ell'}}\cos(\psi_-) \bm{\hat{e}}_{\varphi_q}+\delta_{|\ell-\ell'|,2}\bm{\hat{e}}_{\varphi_q+\text{sgn}(\ell-\ell')\psi_-} \nonumber\\
    & \quad+\delta_{\ell+\ell',2}\bm{\hat{e}}_{\varphi_q+\psi_+}]
    \}.
\end{align}

\noindent {\color{black}However, it remains slowly modulated on the phonon timescale through $\sin(2b_p)$ and $\cos(2b_p)$. Hence it is quasi-dc rather than strictly dc, except in the limit $\Omega_p\rightarrow 0$.} At the resonant condition where the spin and phonon drives have the same time windows, $\Omega_p= \Omega_B$, the time average both spin and phonon drives of this second-correction at $\mathcal{O}(q)$ order vanishes.

{\color{black}
Here we estimate the order of magnitude of the unit-cell-averaged emergent field in the long-wavelength regime ($\lambda\gg a$) for an ideal periodic SkX (no boundary/edge contribution) using the narrow-wall approximation. For $R_0=50$ nm and $A_\text{cell}=3.46\times10^{-14} \;\text{m}^2$, we obtain  $C_f^{(0)}\simeq-3.0\times10^{-9}$~Vs/m, $C_f^{(q)}\simeq-1.5q \times10^{-16}$~Vs/m, and $C_\mathcal{F}^{(q)}\simeq+1.5q \times10^{-16}$ Vs/m. As a representative acoustic drive, we consider a coherent wave with frequency $f_p\sim 1.5$ GHz and wavelength $\lambda\sim 2 \;\mu$m~\cite{PhysRevB.86.134415}, and assume it induces an elliptical deformation $\eta_2=\eta_{02}\cos(2\pi f_B t)$ with $\eta_{02}=0.02$ and $f_B=f_p$. The leading $\mathcal{O}(\eta)$ macroscopic-field amplitude is then $|\langle E_1^e \rangle| = 0.44$ V/m with an oscillating direction is along $\bm{\hat{e}}_{2\varphi_2-\varphi_q}$. Meanwhile, the $\mathcal{O}(\eta^2)$ correction arising from harmonic mixing linearized in $q$-term is parametrically smaller and is estimated to be $10^{-4}-10^{-3} \; $V/m.


}
\subsection{\color{black} Rotational Modes}
\bo{\color{black} Angular-harmonic deformation}. {\color{black}Next, we consider rotational CW/CCW excitation modes in the presence of an angular-harmonic deformation of the skyrmion boundary. We assume that the skyrmion radius acquires a purely azimuthal modulation}, $R(\varphi)=R_0[1+\mathcal{D}(\varphi)]$, with $\mathcal{D}(\varphi)=\sum_{\ell \geq 0}\eta_{0\ell} \cos(\ell(\varphi-\varphi_\ell))$, %
so that the polar profile is time-independent, $\partial_t\Theta(\rho)=0$ ($\rho=r/R(\varphi)$). While, the azimuthal angle precesses {\color{black}uniformly} in time as $\Phi=v\varphi+\gamma+\Omega_G t$. {\color{black}With this ansatz}, the emergent electric field becomes

\begin{align}
    \bm{E}^e(r,\varphi)=\frac{\hbar}{2e}\frac{\Omega_G}{R(\varphi)}f(\rho) [\bm{\hat{e}}_r- \frac{\partial_\varphi R(\varphi)}{R(\varphi)} \bm{\hat{e}}_\varphi],
    \label{eq:E_rot01}
\end{align}
\noindent with $f(\rho)= \sin \Theta(\rho)\partial_\rho\Theta(\rho)$.

Expand the electric field induced for a small lattice distortion up to the second leading order of $\mathcal{D}$,

\begin{align}
    \bm{E}^e(r,\varphi) &=\frac{\hbar}{2e}\frac{\Omega_G}{R_0}\{[f_0-\mathcal{F}_{1}\mathcal{D}+\mathcal{F}_{2}\mathcal{D}^2 ]\bm{\hat{e}}_r \nonumber \\ 
    & \quad +[-f_0\partial_\varphi\mathcal{D} + (f_0+\mathcal{F}_{1}) \mathcal{D}\partial_\varphi\mathcal{D}] \bm{\hat{e}}_\varphi\}
    \label{eq:E_rot02}
\end{align}


\noindent where $f_0=f(\rho_0)$, the shape factor $\mathcal{F}_1$ and $\mathcal{F}_2$,
\begin{align}
    \mathcal{F}_1(\rho_0)&=f(\rho_0)+\rho_0f'(\rho_0) \\
    \mathcal{F}_2(\rho_0) &= f(\rho_0)+2\rho_0f'(\rho_0)+\frac{1}{2}\rho_0^2f''(\rho_0).
\end{align}

\noindent Here, the electric field has an azimuthal component in addition to the radial one. {\color{black} This electric distribution is shown in Fig. \ref{fig:field_maps}(b)}. However, in the case of an infinite system {\color{black}with no edge term}, the average electric fields over the cell area vanish, $\langle\bm{E}^e\rangle=0$. Since it is subjected to the boundary condition, $f(\rho_0),f'(\rho_0)\rightarrow0$, at both ends $\rho_0=[0,{\color{black}\rho_{ws}}]$. In addition, the radial and azimuthal components are also canceled out.
This implies that, at infinite systems, the only azimuthal dependence of lattice distortion cannot produce a coarse-grained electric field.

{\color{black}In the presence of boundary term}, we can introduce {\color{black}an effective Wigner–Seitz cutoff}, ${\color{black}\rho_{ws}}=\Lambda$. Thus, the nonzero contribution of macroscopic electric field in the first leading order $\mathcal{O}(\eta)$, {\color{black}normalized by the circular Wigner-Seitz} cell area $A_c=\pi(\Lambda R_0)^2$, arises only from the dipolar bias ($\ell=1$),

\begin{align}
   \langle \bm{E}_1^e\rangle =-\frac{\hbar}{2e}\frac{\Omega_G}{R_0}\eta_{01}f(\Lambda) \bm{\hat{e}}_r(\varphi_1),
\end{align}

\noindent while the second-order correction {\color{black}originates} from the nearest-neighboring mixing harmonics,
\begin{align}
\langle \bm{E}_2^e \rangle&=\frac{\hbar}{2e}\frac{\Omega_G}{R_0}\mathcal{A}_1\left[f(\Lambda) + \frac{1}{2}\Lambda f'(\Lambda)\right]\bm{\hat{e}}_r(\varphi_1), \\
\mathcal{A}_1&=\frac{1}{4}\sum_{\ell,\ell'\geq0}\eta_{0\ell}\eta_{0\ell'}\big[\delta_{|\ell-\ell'|,1}\cos(\ell\varphi_\ell-\ell'\varphi_{\ell'}-\varphi_1) \nonumber \\
&\quad+\delta_{\ell+\ell',1}\cos(\ell\varphi_\ell+\ell'\varphi_{\ell'}-\varphi_1)\big].
\end{align}

\noindent Both first and second order give a finite macroscopic dc field set solely by the boundary values $f(\Lambda), f'(\Lambda)$. These values vanish for {\color{black} an ideal periodic SkX with no edge term}.

{\color{black}In the case of narrow wall with cutoff radius $\Lambda\rightarrow 1$ (near wall), using the same parameters ($R_0$, $\Delta_w$, and $\eta_{01}$) that have been used for breathing mode, taking the resonance frequency $\Omega_G=2\pi\times1\;$GHz, the magnitude of unit-cell averaged emergent field for the first leading order $\mathcal{O}(\eta)$ is estimated as $|\langle \bm{E}_1^e\rangle|\simeq13$ V/m. Meanwhile, the $O(\eta^2)$ correction arising from nearest-neighbor harmonic mixing is $|\langle \bm{E}_2^e\rangle|\simeq 0.1 \;$V/m. } 



\bo{Phonon coupling}. Now consider the coupling of lattice distortion with the phonon mode. The {\color{black}angular harmonic deformation $\mathcal{D}$} is modified as follows,

\begin{align}
    \mathcal{D}(\bm{r},\varphi,t)=\sum_{\ell \geq 0}\eta_{0\ell} \cos(\ell(\varphi-\varphi_\ell))\cos(\bm{q}\cdot\bm{r}-\Omega_p t+\chi_\ell),
\end{align}


\noindent the induced electric field is modified as,

\begin{align}
    \bm{E}^e(\bm{r},\varphi,t)=\frac{\hbar\Omega_G}{2e}\frac{f(\rho)}{R(\varphi,r,t)} \left[(1-\frac{r\partial_r {\color{black} R} }{\color{black}R})\bm{\hat{e}}_r - 
    \frac{\partial_\varphi {\color{black} R}}{\color{black}R} \bm{\hat{e}}_\varphi\right].
\end{align}
Here, we have additional contribution, $\propto r\partial_r\mathcal{D}$, in the radial direction $\bm{\hat{e}}_r$. Other contributions, as mentioned earlier, leads to zero macroscopic field at infinite systems, except if the boundary effect is considered.    

Expand the contribution from $r\partial_r\mathcal{D}$ term at a small lattice distortion, and use the following vector projection,
\begin{align}
    \bm{\hat{e}}_r(\varphi)=\cos(\theta)\bm{\hat{e}}_r(\varphi_q)+\sin(\theta)\bm{\hat{e}}_\varphi(\varphi_q),
\end{align}
\noindent after some algebra, we find a nonzero macroscopic contribution of the cell-average electric field {\color{black}around the wall region } coming from the first leading order in $\eta$ that couples $\rho_0R_0\partial_r\mathcal{D}|_{r=R_0}$ term,

\begin{align}
    \langle \bm{E}^e_1 \rangle(t) &= \mathcal{B}_q
     \sum_{\ell\geq 0} \eta_{0\ell} [\cos(\ell\Delta_\ell) T_\ell \;\bm{\hat{e}}_r(\varphi_q) \nonumber \\ 
     & \quad +\sin(\ell\Delta_\ell)I_\ell \bm{\hat{e}}_\varphi(\varphi_q)],
\end{align}

\noindent where $\mathcal{B}_q$, $T_\ell$, and $I_\ell$ are defined as 
\begin{align}
\mathcal{B}_q &= \frac{\pi a_q\hbar \Omega_G {\color{black}R_0}}{e \color{black}A_\text{cell}} \int_0^{\color{black}\rho_{ws}} \rho_0^2f(\rho_0) d\rho_0 \\
T_\ell&=  \langle \cos^2 (\theta) \cos(\ell\theta) \sin(a_q \cos\theta+b_\ell)\rangle_\theta \nonumber \\
I_\ell &=\langle {\sin(\theta)\color{black} \cos(\theta) \sin(\ell\theta)}\sin(a_q \cos\theta+b_\ell)  \rangle_\theta
\end{align}

\noindent with {\color{black}the variables} $\theta=\varphi-\varphi_q$, $a_q=qR_0$, $b_\ell = -\Omega_p t+\chi_\ell$, and $\Delta_\ell=\varphi_\ell-\varphi_q$. In the following, we write the values of $T_\ell$ and $I_\ell$ at the long-wavelength limit ($qR_0\ll 1$) up to $\ell=4$ and $\mathcal{O}(a^7)$,

\begin{align}
    T_0 &= \sin(b)\left[\frac{1}{2}-\frac{3a^2}{16} + \frac{5a^4}{384} \right] + \mathcal{O}(a^6)\\
    I_0&= \sin(b)\left[\frac{1}{2}-\frac{a^2}{16}+\frac{a^4}{384}\right] + \mathcal{O}(a^6)\\
    T_1&=\cos (b) \left[\frac{3a}{8}-\frac{5a^3}{96}+\frac{7a^5}{3072}\right]+\mathcal{O}(a^7), \\
    I_1 &= \frac{1}{8} \cos(b) \left[a-\frac{a^3}{12}+\frac{a^5}{384}\right]+\mathcal{O}(a^7), \\
    T_2&=\sin(b)\left[ \frac{1}{4} - \frac{a^2}{8} +\frac{5a^4}{512} \right] +\mathcal{O}(a^6), \\
    I_2 &= \frac{1}{4} \sin(b) \left[1-\frac{a^2}{4}+\frac{5a^4}{384}\right]+\mathcal{O}(a^6), \\
    T_3 &=\cos(b)\left[\frac{a}{8}-\frac{5a^3}{192}+\frac{7a^5}{5120} \right] +\mathcal{O}(a^7), \\
    I_3 &= \frac{1}{8} \cos(b) \left[a-\frac{a^3}{8}+\frac{3a^5}{640}\right]+\mathcal{O}(a^7), \\
    T_4 &=\sin(b)\left[ -\frac{a^2}{32}+\frac{a^4}{256} \right]  + \mathcal{O}(a^6),\\
    I_4 &= \frac{1}{8} \sin(b) \left[-\frac{a^2}{4}+\frac{a^4}{48}\right]+\mathcal{O}(a^6).
\end{align}

\noindent {We find that the dominant contribution at the long-wavelength limit arises from phonon-coupled isotropic ($\ell=0$) and elliptical ($\ell=2$) deformation, the macroscopic oscillating electric field is linearly scaled by momentum, $a_q=qR_0$. While in the dipolar ($\ell=1$) and and triangular ($\ell=3$) bias, they are scaled by $(qR_0)^2$. However, note that this first-order expansion in the electric field does not produce the macroscopic dc field since it has a factor of $\sin(b_\ell)$ or $\cos(b_\ell)$, in which the time average of this factor is zero, $\overline{\langle E_1^e \rangle}^t = 0$.

{\color{black}
For representative values \(R_0=50~\mathrm{nm}\), \(\Omega_G=2\pi\times 1~\mathrm{GHz}\),
\(\lambda_p=2~\mu\mathrm{m}\), and \(A_{\rm cell}\sim 10^{-14}~\mathrm{m}^2\), with the narrow-wall estimation
\(\int_0^{\rho_{\rm ws}}\rho_0^2 f(\rho_0)\,d\rho_0 \simeq -2\),
we obtain a numerical value of \(|B_q| \simeq 20.5~\mathrm{V/m}\). Assuming isotropic deformation with amplitude $\eta_{00}\approx 0.02$, the unit-cell averaged electric field is estimated as $0.41$ V/m.}

Let us examine the second-order correction in the electric field,

\begin{align}
    \bm{E}_2^e(\bm{r},\varphi,t) &=\frac{\hbar \Omega_G}{2e R_0}\big[(f+\mathcal{F}_1)(\mathcal{D}\partial_\varphi\mathcal{D}\bm{\hat{e}}_\varphi+\rho_0R_0\mathcal{D}\partial_r\mathcal{D}\bm{\hat{e}}_r) \nonumber \\ 
    & \qquad \qquad+ \mathcal{F}_2\mathcal{D}^2\bm{\hat{e}}_r
    \big].
\end{align}


\noindent The non-zero contribution of the coarse-grained field only comes from the $\partial_r\mathcal{D}$ term,

\begin{align}
    \bm{E}_2^e(\bm{r},\varphi,t) &=\frac{\hbar \Omega_G}{2e}\rho_0(2f+\rho_0f')\mathcal{D}\partial_r\mathcal{D} \bm{\hat{e}}_r
\end{align}
\noindent $\mathcal{D}\partial_r\mathcal{D}$ term can be expanded around $r=R_0$,
\begin{align}
    \mathcal{D}\partial_r\mathcal{D}&=\frac{1}{R_0}\sum_{n,m\geq0} \frac{(x-1)^{n+m}}{n!m!} \mathcal{D}_n(\varphi,t)\mathcal{D}_{m+1}(\varphi,t) \\
    \mathcal{D}_n(\varphi,t)&=\partial_x^n\mathcal{D}(\varphi,r,t)|_{x=1}, \qquad x=\rho_0=\frac{r}{R_0}
\end{align}

\noindent Thus, the cell average of the second-order electric field,

\begin{align}
    \langle\bm{E}_2^e(\bm{r},\varphi,t)\rangle &=\frac{\hbar\Omega_G {\color{black} R_0}}{2e \color{black} A_\text{cell}} \sum_{n,m\geq 0}\frac{J_{n+m}}{n!m!} \nonumber \\
    & \quad \times\int_0^{2\pi}\mathcal{D}_n(\varphi,t)\mathcal{D}_{m+1}(\varphi,t) \bm{\hat{e}}_r(\varphi)d\varphi 
\end{align}

\noindent where,
\begin{align}
    J_{n+m}=\int_0^{\color{black}\rho_{ws} }x^2(x-1)^{n+m}(2f+xf')dx.
\end{align}

\noindent The derivative of $\mathcal{D}_n(\varphi,t)$ can be expressed as

\begin{align}
\mathcal{D}_n(\varphi,t)&=\sum_{\ell} \eta_{0\ell} \cos(\ell(\varphi-\varphi_\ell))(a_q\cos(\theta))^n \nonumber\\ 
&\quad \times\cos(a_q\cos (\theta)+b_\ell+n\pi/2).    
\end{align}

\noindent The product of $\mathcal{D}_n\mathcal{D}_{m+1}$ carries a factor of $(a_q\cos(\theta))^{n+m+1}$. Here, we are concerned in the long-wavelength limit with the smallest power in $a_q=qR_0\ll 1$, where the parameter $n=0, m=0$, or the product of $\mathcal{D}_0\mathcal{D}_1$. {\color{black}Assuming phonon coupling is phase-locked $b_\ell=b_{\ell'}=b_p$}. The sum of higher-order contributions of $a_q^l,l>1$ in $\sum_{n,m}\mathcal{D}_n\mathcal{D}_{m+1}$ vanishes.  
\begin{align}
    \mathcal{D}_0\mathcal{D}_1&=-\frac{a_q}{2}\sin(2b_p)\cos(\theta)\sum_{\ell,\ell'} \eta_{0\ell}\eta_{0\ell'} \cos(\ell(\varphi-\varphi_\ell)) \nonumber \\
    &\quad \times\cos(\ell'(\varphi-\varphi_{\ell'})) + \mathcal{O}(a_q^2).
\end{align}

\noindent The vector projector is given by
\begin{align}
    \bm{\hat{e}}_r(\varphi)=\cos(\theta)\bm{\hat{e}}_r(\varphi_q)+\sin(\theta)\bm{\hat{e}}_\varphi(\varphi_q)
\end{align}

\noindent with the $\sin(\theta)$ branch averages to zero, while the $\cos(\theta)$ branch survives. Suppose that $\varphi_\ell=\varphi_{\ell'}=\varphi_d$ and $\Delta_d=\varphi_d-\varphi_q$. After some algebra, the second-order cell average of the macroscopic electric field linearized in $q$ leads to
\begin{align}
\langle\bm{E}_2^e\rangle(t)&=\mathcal{A}^{(q)}_0\sin(2b_p) \left[  \frac{1}{4} \cos(2\Delta_d)\eta_{01}^2 + \sum_\ell\mathcal{C}_\ell\right]\bm{\hat{e}}_r(\varphi_q),  
\end{align}

\noindent where $\mathcal{A}_0^{(q)}$ and $\mathcal{C}_\ell$ are defined as
\begin{align}
    \mathcal{A}_0^{(q)} &= -\frac{\pi a_q \hbar\Omega_G {\color{black} R_0}}{4e \color{black}A_\text{cell}}J_0, \\
\mathcal{C}_\ell &= \frac{1}{2}\eta_{0\ell}^2 + \frac{1}{2}\cos(2\Delta_d)\eta_{0\ell}\eta_{0,\ell+2}.
\end{align}


\begin{table*}[t]
\caption{\color{black}Selection rules for the non-zero unit-cell averaged emergent electric field
$\langle \mathbf{E}^e\rangle$ for uncoupled deformation ($q=0$) and phonon-coupled deformation ($q\neq0$), keeping the first and second
orders in the deformation amplitude $\eta$.
}
\label{tab:selection_rules_Eavg}
\begin{ruledtabular}
\begin{tabular}{lcccc}
& \multicolumn{2}{c}{Harmonic deformation (no phonon, $q=0$)} &
\multicolumn{2}{c}{Phonon-coupled ($q\neq0$, leading in $q$)} \\
\cline{2-5} 
Mode &
$\mathcal{O}(\eta)$ &
$\mathcal{O}(\eta^2)$ &
$\mathcal{O}(\eta)$ &
$\mathcal{O}(\eta^2)$ \\ [3pt]
\hline \\ [0.5pt]
Breathing &  dipolar$^1$ & AHM$^{1,3,*}$ & isotropic, elliptical$^{1}$ & DHM, NHM$^{1,3}$ \\ [6pt]
Rotational &
dipolar$^{2,*}$ &   AHM$^{2,*}$ & isotropic, elliptical$^{1}$ & DHM, NHM$^{1,3}$   \\ [6pt]
\cline{1-5}
AHM: adjacent harmonic mixing\\
DHM: diagonal harmonic mixing\\
NHM: next-nearest harmonic mixing\\ 
$^1$Oscillating ac field\\
$^2$Rectified dc field\\
$^3$Rect. dc field owing to finite phase lags\\ 
$^*$Requires an edge/boundary term\\
\end{tabular}
\end{ruledtabular}

\end{table*}

\noindent{\color{black} By using the same parameters as in the first-order expansion, the numerical value of $\mathcal{A}_0^{(q)}$ is estimated as $-2.55J_0\approx5.1$ V/m. Therefore, the second-order correction $\mathcal{O}(\eta^2)$ of the unit-cell-averaged electric field is on the order of $10^{-4}-10^{-3}$ V/m.}

This result implies that the second-order correction originates from the diagonal component ($\eta_{0\ell}^2$) and second-neighbor harmonic mixing ($\eta_{0\ell}\eta_{0,\ell+2}$). The resulting field is oscillating along the phonon propagation $\bm{\hat{e}}_r(\varphi_q)$. This second-order correction is proportional to the oscillating phonon, $\propto \sin(2(-\Omega_pt_p+\chi))$. Consequently, the time average of this result vanishes. Thus, the macroscopic dc field is absent. However, if the $\ell$-th harmonic distortion has a different phonon phase factor of $\chi_\ell$, we find that the macroscopic dc field does not vanish, which originates from the second-nearest harmonic mixing,

 \begin{align}
\overline{\langle\bm{E}_2^e\rangle}^{t_p}&= \frac{\mathcal{A}_0^{(q)}}{2}\cos(2\Delta_d)\sum_\ell \eta_{0\ell}\eta_{0,\ell+2}\sin(\chi_{\ell+2}-\chi_\ell)\bm{\hat{e}}_r(\varphi_q), 
\end{align}

\noindent the response is maximized when the two harmonic components are in quadrature ($\Delta \chi=\pi/2$), and vanishes for in-phase ($\Delta \chi=0$) or antiphase ($\Delta \chi=\pi$) conditions.

\section{Summary}
In summary, we have derived the macroscopic electric field generated by a small lattice perturbations arising from the phonon-coupled lattice deformations in the internal breathing and rotational dynamics of a skyrmion lattice under microwave excitations. By coarse-graining the Berry-phase electric field over a skyrmion-lattice unit cell, we identify the symmetry and dynamical conditions under which the collective spin-lattice motion yields a finite macroscopic response{\color{black}, as provided in Table \ref{tab:selection_rules_Eavg}}. 

Our analysis clarifies that both rectified dc and oscillating electric fields can emerge when the breathing and rotational modes are coupled to lattice strain with broken inversion or rotational symmetry. With experimentally determined skyrmion profile parameters such as the equilibrium radius, domain-wall width, and resonance frequency, our theoretical framework can be used to identify the specific harmonic components that contribute to the measured macroscopic electric field. 

These results may provide a unified microscopic basis for phonon-driven electrodynamics in topological spin textures, highlighting a pathway toward dynamical spin–charge–lattice interconversion in skyrmion crystals.

\begin{center}
\bo{ACKNOWLEDGMENTS}
\end{center}

This theoretical study was supported by PUTI Q1 Research Grant No. NKB-438/UN2.RST/HKP.05.00/2024, from Universitas Indonesia, in the period of 2024/2025. 

\bibliographystyle{apsrev4-2}
\bibliography{reference}

@Article{Nagaosa2013_NatNano,
author={Nagaosa, Naoto
and Tokura, Yoshinori},
title={Topological properties and dynamics of magnetic skyrmions},
journal={Nature Nanotechnology},
year={2013},
month={Dec},
day={01},
volume={8},
number={12},
pages={899-911},
issn={1748-3395},
doi={10.1038/nnano.2013.243},
url={https://doi.org/10.1038/nnano.2013.243}
}

@Article{Fert2017_NatRevMat,
author={Fert, Albert
and Reyren, Nicolas
and Cros, Vincent},
title={Magnetic skyrmions: advances in physics and potential applications},
journal={Nature Reviews Materials},
year={2017},
month={Jun},
day={13},
volume={2},
number={7},
pages={17031},
issn={2058-8437},
doi={10.1038/natrevmats.2017.31},
url={https://doi.org/10.1038/natrevmats.2017.31}
}

@Article{Schulz2012_NatPhys,
author={Schulz, T.
and Ritz, R.
and Bauer, A.
and Halder, M.
and Wagner, M.
and Franz, C.
and Pfleiderer, C.
and Everschor, K.
and Garst, M.
and Rosch, A.},
title={Emergent electrodynamics of skyrmions in a chiral magnet},
journal={Nature Physics},
year={2012},
month={Apr},
day={01},
volume={8},
number={4},
pages={301-304},
issn={1745-2481},
doi={10.1038/nphys2231},
url={https://doi.org/10.1038/nphys2231}
}

@article{Rosch2013_NatNano,
  author    = {Achim Rosch},
  title     = {Skyrmions: Moving with the current},
  journal   = {Nature Nanotechnology},
  volume    = {8},
  number    = {3},
  pages     = {160--161},
  year      = {2013},
  doi       = {10.1038/nnano.2013.21}
}

@article{Li2023_IDM2,
author = {Li, Sheng and Wang, Xuewen and Rasing, Theo},
title = {Magnetic skyrmions: Basic properties and potential applications},
journal = {Interdisciplinary Materials},
volume = {2},
number = {2},
pages = {260-289},
doi = {https://doi.org/10.1002/idm2.12072},
year = {2023}
}

@article{Neubauer2009_PRL,
  title = {Topological Hall Effect in the $A$ Phase of MnSi},
  author = {Neubauer, A. and Pfleiderer, C. and Binz, B. and Rosch, A. and Ritz, R. and Niklowitz, P. G. and B\"oni, P.},
  journal = {Phys. Rev. Lett.},
  volume = {102},
  issue = {18},
  pages = {186602},
  numpages = {4},
  year = {2009},
  month = {May},
  publisher = {American Physical Society},
  doi = {10.1103/PhysRevLett.102.186602},
  url = {https://link.aps.org/doi/10.1103/PhysRevLett.102.186602}
}

@article{Tatara_PhysRevLett.92.086601,
  title = {Theory of Current-Driven Domain Wall Motion: Spin Transfer versus Momentum Transfer},
  author = {Tatara, Gen and Kohno, Hiroshi},
  journal = {Phys. Rev. Lett.},
  volume = {92},
  issue = {8},
  pages = {086601},
  numpages = {4},
  year = {2004},
  month = {Feb},
  publisher = {American Physical Society},
  doi = {10.1103/PhysRevLett.92.086601},
  url = {https://link.aps.org/doi/10.1103/PhysRevLett.92.086601}
}

@article{Tokura2021_ChemRev,
  author    = {Y. Tokura and N. Kanazawa},
  title     = {Magnetic Skyrmion Materials},
  journal   = {Chemical Reviews},
  volume    = {121},
  number    = {5},
  pages     = {2857--2897},
  year      = {2021},
  doi       = {10.1021/acs.chemrev.0c00297}
}

@article{EverschorSitte2018_JAP,
  author    = {K. Everschor-Sitte and J. Masell and R. M. Reeve and M. Kläui},
  title     = {Perspective: Magnetic skyrmions—Overview of recent progress in an active research field},
  journal   = {Journal of Applied Physics},
  volume    = {124},
  number    = {24},
  pages     = {240901},
  year      = {2018},
  doi       = {10.1063/1.5048972}
}

@article{Lin2016_PRB,
  title = {Dynamics of Dirac strings and monopolelike excitations in chiral magnets under a current drive},
  author = {Lin, Shi-Zeng and Saxena, Avadh},
  journal = {Phys. Rev. B},
  volume = {93},
  issue = {6},
  pages = {060401},
  numpages = {5},
  year = {2016},
  month = {Feb},
  publisher = {American Physical Society},
  doi = {10.1103/PhysRevB.93.060401},
  url = {https://link.aps.org/doi/10.1103/PhysRevB.93.060401}
}

@article{Koide_PhysRevB.100.014408,
  title = {DC spinmotive force from microwave-active resonant dynamics of a skyrmion crystal under a tilted magnetic field},
  author = {Koide, Tatsuya and Takeuchi, Akihito and Mochizuki, Masahito},
  journal = {Phys. Rev. B},
  volume = {100},
  issue = {1},
  pages = {014408},
  numpages = {9},
  year = {2019},
  month = {Jul},
  publisher = {American Physical Society},
  doi = {10.1103/PhysRevB.100.014408},
  url = {https://link.aps.org/doi/10.1103/PhysRevB.100.014408}
}

@article{Garst2017_JPhysD,
  author    = {M. Garst and J. Waizner and D. Grundler},
  title     = {Collective spin excitations of helices and magnetic skyrmions: review and perspectives of magnonics in non-centrosymmetric magnets},
  journal   = {Journal of Physics D: Applied Physics},
  volume    = {50},
  pages     = {293002},
  year      = {2017},
  doi       = {10.1088/1361-6463/aa7573}
}

@article{Mochizuki2012_PRL,
  author    = {M. Mochizuki},
  title     = {Spin-Wave Modes and Their Intense Excitation Effects in Skyrmion Crystals},
  journal   = {Physical Review Letters},
  volume    = {108},
  pages     = {017601},
  year      = {2012},
  doi       = {10.1103/PhysRevLett.108.017601}
}

@article{Iwasaki2013_NatCommun,
  author    = {J. Iwasaki and M. Mochizuki and N. Nagaosa},
  title     = {Universal current-velocity relation of skyrmion motion in chiral magnets},
  journal   = {Nature Communications},
  volume    = {4},
  pages     = {1463},
  year      = {2013},
  doi       = {10.1038/ncomms2442}
}

@article{Mochizuki2014_NatMater,
  author    = {M. Mochizuki and X. Z. Yu and S. Seki and N. Kanazawa and W. Koshibae and J. Zang and M. Mostovoy and Y. Tokura and N. Nagaosa},
  title     = {Thermally driven ratchet motion of a skyrmion microcrystal and topological magnon Hall effect},
  journal   = {Nature Materials},
  volume    = {13},
  pages     = {241--246},
  year      = {2014},
  doi       = {10.1038/nmat3862}
}

@article{Onose2012_PRL,
  author    = {Y. Onose and Y. Okamura and S. Seki and S. Ishiwata and Y. Tokura},
  title     = {Observation of Magnetic Excitations of Skyrmion Crystal in a Helimagnetic Insulator Cu$_2$OSeO$_3$},
  journal   = {Physical Review Letters},
  volume    = {109},
  pages     = {037603},
  year      = {2012},
  doi       = {10.1103/PhysRevLett.109.037603}
}

@Article{Okamura2013_NatCommun,
author={Okamura, Y.
and Kagawa, F.
and Mochizuki, M.
and Kubota, M.
and Seki, S.
and Ishiwata, S.
and Kawasaki, M.
and Onose, Y.
and Tokura, Y.},
title={Microwave magnetoelectric effect via skyrmion resonance modes in a helimagnetic multiferroic},
journal={Nature Communications},
year={2013},
month={Aug},
day={30},
volume={4},
number={1},
pages={2391},
issn={2041-1723},
doi={10.1038/ncomms3391},
url={https://doi.org/10.1038/ncomms3391}
}

@article{Lucassen_PhysRevB.84.014414,
  title = {Spin motive forces due to magnetic vortices and domain walls},
  author = {Lucassen, M. E. and Kruis, G. C. F. L. and Lavrijsen, R. and Swagten, H. J. M. and Koopmans, B. and Duine, R. A.},
  journal = {Phys. Rev. B},
  volume = {84},
  issue = {1},
  pages = {014414},
  numpages = {7},
  year = {2011},
  month = {Jul},
  publisher = {American Physical Society},
  doi = {10.1103/PhysRevB.84.014414},
  url = {https://link.aps.org/doi/10.1103/PhysRevB.84.014414}
}

@article{Lin2015_PRB,
  title = {Noncircular skyrmion and its anisotropic response in thin films of chiral magnets under a tilted magnetic field},
  author = {Lin, Shi-Zeng and Saxena, Avadh},
  journal = {Phys. Rev. B},
  volume = {92},
  issue = {18},
  pages = {180401},
  numpages = {5},
  year = {2015},
  month = {Nov},
  publisher = {American Physical Society},
  doi = {10.1103/PhysRevB.92.180401},
  url = {https://link.aps.org/doi/10.1103/PhysRevB.92.180401}
}

@article{Masell_PhysRevB.101.214428,
  title = {Spin-transfer torque driven motion, deformation, and instabilities of magnetic skyrmions at high currents},
  author = {Masell, J. and Rodrigues, D. R. and McKeever, B. F. and Everschor-Sitte, K.},
  journal = {Phys. Rev. B},
  volume = {101},
  issue = {21},
  pages = {214428},
  numpages = {13},
  year = {2020},
  month = {Jun},
  publisher = {American Physical Society},
  doi = {10.1103/PhysRevB.101.214428},
  url = {https://link.aps.org/doi/10.1103/PhysRevB.101.214428}
}

@article{Hu2019_PRB_214412,
  author  = {Yangfan Hu and Xiaoming Lan and Biao Wang},
  title   = {Nonlinear emergent elasticity and structural transitions of a skyrmion crystal under uniaxial distortion},
  journal = {Physical Review B},
  volume  = {99},
  pages   = {214412},
  year    = {2019},
  doi     = {10.1103/PhysRevB.99.214412}
}

@article{Hu2019_PRB_144424,
  author  = {Yangfan Hu},
  title   = {Long-wavelength emergent phonons in Bloch skyrmion crystals distorted by exchange anisotropy and tilted magnetic fields},
  journal = {Physical Review B},
  volume  = {100},
  pages   = {144424},
  year    = {2019},
  doi     = {10.1103/PhysRevB.100.144424}
}

@article{Adachi_PhysRevB.109.144413,
  title = {Elastic and anelastic behavior associated with magnetic ordering in the skyrmion host ${\text{Cu}}_{2}{\text{OSeO}}_{3}$},
  author = {Adachi, Kanta and Wilhelm, Heribert and Schmidt, Marcus P. and Carpenter, Michael A.},
  journal = {Phys. Rev. B},
  volume = {109},
  issue = {14},
  pages = {144413},
  numpages = {10},
  year = {2024},
  month = {Apr},
  publisher = {American Physical Society},
  doi = {10.1103/PhysRevB.109.144413},
  url = {https://link.aps.org/doi/10.1103/PhysRevB.109.144413}
}

@article{Petrova2011_PRB,
  author    = {Olga Petrova and Oleg Tchernyshyov},
  title     = {Spin waves in a skyrmion crystal},
  journal   = {Physical Review B},
  volume    = {84},
  pages     = {214433},
  year      = {2011},
  doi       = {10.1103/PhysRevB.84.214433}
}

@article{Hu_2017,
doi = {10.1088/1367-2630/aa9507},
url = {https://doi.org/10.1088/1367-2630/aa9507},
year = {2017},
month = {dec},
publisher = {IOP Publishing},
volume = {19},
number = {12},
pages = {123002},
author = {Hu, Yangfan and Wang, Biao},
title = {Unified theory of magnetoelastic effects in B20 chiral magnets},
journal = {New Journal of Physics}
}

@article{Spachmann2021_PRB_103,
  author    = {S. Spachmann and A. Elghandour and M. Frontzek and W. L{\"o}ser and R. Klingeler},
  title     = {Magnetoelastic coupling and phases in the skyrmion-lattice magnet},
  journal   = {Physical Review B},
  volume    = {103},
  pages     = {184424},
  year      = {2021},
  doi       = {10.1103/PhysRevB.103.184424}
}

@Article{Nii2015_NatCommun,
author={Nii, Y.
and Nakajima, T.
and Kikkawa, A.
and Yamasaki, Y.
and Ohishi, K.
and Suzuki, J.
and Taguchi, Y.
and Arima, T.
and Tokura, Y.
and Iwasa, Y.},
title={Uniaxial stress control of skyrmion phase},
journal={Nature Communications},
year={2015},
month={Oct},
day={13},
volume={6},
number={1},
pages={8539},
issn={2041-1723},
doi={10.1038/ncomms9539},
url={https://doi.org/10.1038/ncomms9539}
}

@article{Nomura_PhysRevLett.122.145901,
  title = {Phonon Magnetochiral Effect},
  author = {Nomura, T. and Zhang, X.-X. and Zherlitsyn, S. and Wosnitza, J. and Tokura, Y. and Nagaosa, N. and Seki, S.},
  journal = {Phys. Rev. Lett.},
  volume = {122},
  issue = {14},
  pages = {145901},
  numpages = {5},
  year = {2019},
  month = {Apr},
  publisher = {American Physical Society},
  doi = {10.1103/PhysRevLett.122.145901},
  url = {https://link.aps.org/doi/10.1103/PhysRevLett.122.145901}
}

@article{Zhang_PhysRevLett.102.086601,
  title = {Generalization of the Landau-Lifshitz-Gilbert Equation for Conducting Ferromagnets},
  author = {Zhang, Shufeng and Zhang, Steven S.-L.},
  journal = {Phys. Rev. Lett.},
  volume = {102},
  issue = {8},
  pages = {086601},
  numpages = {4},
  year = {2009},
  month = {Feb},
  publisher = {American Physical Society},
  doi = {10.1103/PhysRevLett.102.086601},
  url = {https://link.aps.org/doi/10.1103/PhysRevLett.102.086601}
}

@article{Everschor-Sitte_10.1063/1.4870695,
    author = {Everschor-Sitte, Karin and Sitte, Matthias},
    title = {Real-space Berry phases: Skyrmion soccer (invited)},
    journal = {Journal of Applied Physics},
    volume = {115},
    number = {17},
    pages = {172602},
    year = {2014},
    month = {04},
    issn = {0021-8979},
    doi = {10.1063/1.4870695},
    url = {https://doi.org/10.1063/1.4870695},
}

@article{Lu_PhysRevB.91.100405,
  title = {General microscopic model of magnetoelastic coupling from first principles},
  author = {Lu, X. Z. and Wu, Xifan and Xiang, H. J.},
  journal = {Phys. Rev. B},
  volume = {91},
  issue = {10},
  pages = {100405},
  numpages = {5},
  year = {2015},
  month = {Mar},
  publisher = {American Physical Society},
  doi = {10.1103/PhysRevB.91.100405},
  url = {https://link.aps.org/doi/10.1103/PhysRevB.91.100405}
}

@article{Callen_PhysRev.129.578,
  title = {Static Magnetoelastic Coupling in Cubic Crystals},
  author = {Callen, Earl R. and Callen, Herbert B.},
  journal = {Phys. Rev.},
  volume = {129},
  issue = {2},
  pages = {578--593},
  numpages = {0},
  year = {1963},
  month = {Jan},
  publisher = {American Physical Society},
  doi = {10.1103/PhysRev.129.578},
  url = {https://link.aps.org/doi/10.1103/PhysRev.129.578}
}

@article{Keever_PhysRevB.99.054430,
  title = {Characterizing breathing dynamics of magnetic skyrmions and antiskyrmions within the Hamiltonian formalism},
  author = {McKeever, B. F. and Rodrigues, D. R. and Pinna, D. and Abanov, Ar. and Sinova, Jairo and Everschor-Sitte, K.},
  journal = {Phys. Rev. B},
  volume = {99},
  issue = {5},
  pages = {054430},
  numpages = {15},
  year = {2019},
  month = {Feb},
  publisher = {American Physical Society},
  doi = {10.1103/PhysRevB.99.054430},
  url = {https://link.aps.org/doi/10.1103/PhysRevB.99.054430}
}

@article{Verma2022_SciAdv,
  author    = {Nishchhal Verma and Zachariah Addison and Mohit Randeria},
  title     = {Unified theory of the anomalous and topological Hall effects with phase-space Berry curvatures},
  journal   = {Science Advances},
  volume    = {8},
  number    = {45},
  pages     = {eabq2765},
  year      = {2022},
  doi       = {10.1126/sciadv.abq2765}
}

@article{Xiao_RevModPhys.82.1959,
  title = {Berry phase effects on electronic properties},
  author = {Xiao, Di and Chang, Ming-Che and Niu, Qian},
  journal = {Rev. Mod. Phys.},
  volume = {82},
  issue = {3},
  pages = {1959--2007},
  numpages = {0},
  year = {2010},
  month = {Jul},
  publisher = {American Physical Society},
  doi = {10.1103/RevModPhys.82.1959},
  url = {https://link.aps.org/doi/10.1103/RevModPhys.82.1959}
}

@article{Zhang_PhysRevB.101.024420,
  title = {Real-space Berry curvature of itinerant electron systems with spin-orbit interaction},
  author = {Zhang, Shang-Shun and Ishizuka, Hiroaki and Zhang, Hao and Hal\'asz, G\'abor B. and Batista, Cristian D.},
  journal = {Phys. Rev. B},
  volume = {101},
  issue = {2},
  pages = {024420},
  numpages = {15},
  year = {2020},
  month = {Jan},
  publisher = {American Physical Society},
  doi = {10.1103/PhysRevB.101.024420},
  url = {https://link.aps.org/doi/10.1103/PhysRevB.101.024420}
}

@article{Lin_PhysRevB.96.014407,
  title = {Dynamics and inertia of a skyrmion in chiral magnets and interfaces: A linear response approach based on magnon excitations},
  author = {Lin, Shi-Zeng},
  journal = {Phys. Rev. B},
  volume = {96},
  issue = {1},
  pages = {014407},
  numpages = {10},
  year = {2017},
  month = {Jul},
  publisher = {American Physical Society},
  doi = {10.1103/PhysRevB.96.014407},
  url = {https://link.aps.org/doi/10.1103/PhysRevB.96.014407}
}

@Article{Büttner2015,
author={B{\"u}ttner, Felix
and Moutafis, C.
and Schneider, M.
and Kr{\"u}ger, B.
and G{\"u}nther, C. M.
and Geilhufe, J.
and Schmising, C. v. Korff
and Mohanty, J.
and Pfau, B.
and Schaffert, S.
and Bisig, A.
and Foerster, M.
and Schulz, T.
and Vaz, C. A. F.
and Franken, J. H.
and Swagten, H. J. M.
and Kl{\"a}ui, M.
and Eisebitt, S.},
title={Dynamics and inertia of skyrmionic spin structures},
journal={Nature Physics},
year={2015},
month={Mar},
day={01},
volume={11},
number={3},
pages={225-228},
issn={1745-2481},
doi={10.1038/nphys3234},
url={https://doi.org/10.1038/nphys3234}
}

@Article{Wu2021,
author={Wu, Haitao
and Hu, Xuchong
and Jing, Keyu
and Wang, X. R.},
title={Size and profile of skyrmions in skyrmion crystals},
journal={Communications Physics},
year={2021},
month={Sep},
day={20},
volume={4},
number={1},
pages={210},
issn={2399-3650},
doi={10.1038/s42005-021-00716-y},
url={https://doi.org/10.1038/s42005-021-00716-y}
}

@Article{Meyer2019,
author={Meyer, Sebastian
and Perini, Marco
and von Malottki, Stephan
and Kubetzka, Andr{\'e}
and Wiesendanger, Roland
and von Bergmann, Kirsten
and Heinze, Stefan},
title={Isolated zero field sub-10{\thinspace}nm skyrmions in ultrathin Co films},
journal={Nature Communications},
year={2019},
month={Aug},
day={23},
volume={10},
number={1},
pages={3823},
issn={2041-1723},
doi={10.1038/s41467-019-11831-4},
url={https://doi.org/10.1038/s41467-019-11831-4}
}

@article{Rohart_PhysRevB.88.184422,
  title = {Skyrmion confinement in ultrathin film nanostructures in the presence of Dzyaloshinskii-Moriya interaction},
  author = {Rohart, S. and Thiaville, A.},
  journal = {Phys. Rev. B},
  volume = {88},
  issue = {18},
  pages = {184422},
  numpages = {8},
  year = {2013},
  month = {Nov},
  publisher = {American Physical Society},
  doi = {10.1103/PhysRevB.88.184422},
  url = {https://link.aps.org/doi/10.1103/PhysRevB.88.184422}
}

@Article{Wang2018,
author={Wang, X. S.
and Yuan, H. Y.
and Wang, X. R.},
title={A theory on skyrmion size},
journal={Communications Physics},
year={2018},
month={Jul},
day={04},
volume={1},
number={1},
pages={31},
issn={2399-3650},
doi={10.1038/s42005-018-0029-0},
url={https://doi.org/10.1038/s42005-018-0029-0}
}

@article{Zheng_10.1063/5.0152543,
    author = {Zheng, Shasha and Wang, Zhenyu and Wang, Yipu and Sun, Fengxiao and He, Qiongyi and Yan, Peng and Yuan, H. Y.},
    title = {Tutorial: Nonlinear magnonics},
    journal = {Journal of Applied Physics},
    volume = {134},
    number = {15},
    pages = {151101},
    year = {2023},
    month = {10},
    issn = {0021-8979},
    doi = {10.1063/5.0152543},
    url = {https://doi.org/10.1063/5.0152543},
}

@article{Liu_PhysRevB.110.184413,
  title = {Low-lying magnon frequency comb in skyrmion crystals},
  author = {Liu, Xuejuan and Jin, Zhejunyu and Li, Zhengyi and Zeng, Zhaozhuo and Li, Minghao and Yao, Yuping and Cao, Yunshan and Zhang, Yinghui and Yan, Peng},
  journal = {Phys. Rev. B},
  volume = {110},
  issue = {18},
  pages = {184413},
  numpages = {7},
  year = {2024},
  month = {Nov},
  publisher = {American Physical Society},
  doi = {10.1103/PhysRevB.110.184413},
  url = {https://link.aps.org/doi/10.1103/PhysRevB.110.184413}
}

@Article{Lan2025,
author={Lan, Guibin
and Liu, Kang-Yuan
and Wang, Zhenyu
and Xia, Fan
and Xu, Hongjun
and Guo, Tengyu
and Zhang, Yu
and He, Bin
and Li, Jiahui
and Wan, Caihua
and Bauer, Gerrit E. W.
and Yan, Peng
and Liu, Gang-Qin
and Pan, Xin-Yu
and Han, Xiufeng
and Yu, Guoqiang},
title={Coherent harmonic generation of magnons in spin textures},
journal={Nature Communications},
year={2025},
month={Jan},
day={30},
volume={16},
number={1},
pages={1178},
issn={2041-1723},
doi={10.1038/s41467-025-56558-7},
url={https://doi.org/10.1038/s41467-025-56558-7}
}

@Article{Li2020,
author={Li, Xu
and Lynch, Christopher S.},
title={Strong electric field tuning of magnetism in self-biased multiferroic structures},
journal={Scientific Reports},
year={2020},
month={Dec},
day={03},
volume={10},
number={1},
pages={21148},
issn={2045-2322},
doi={10.1038/s41598-020-78104-9},
url={https://doi.org/10.1038/s41598-020-78104-9}
}

@article{Aji_2025,
doi = {10.1088/1361-6463/adf3c2},
url = {https://doi.org/10.1088/1361-6463/adf3c2},
year = {2025},
month = {aug},
publisher = {IOP Publishing},
volume = {58},
number = {31},
pages = {315003},
author = {Aji, Seno and Anin Nabail Azhiim, Muhammad and Ika Puji Ayu, Nur and Badra Cahaya, Adam and Kusakabe, Koichi and Aziz Majidi, Muhammad},
title = {Spin current generation driven by skyrmion dynamics under magnetic anisotropy and polarized microwaves},
journal = {Journal of Physics D: Applied Physics}
}

@book{LandauLifshitz_TheoryElasticity,
  author    = {Landau, L. D. and Lifshitz, E. M.},
  title     = {Theory of Elasticity},
  series    = {Course of Theoretical Physics},
  volume    = {7},
  edition   = {3},
  publisher = {Butterworth-Heinemann},
  address   = {Oxford},
  year      = {1986},
  isbn      = {978-0-7506-2633-0}
}

@misc{Kuhl_ContinuumMechanicsNotes,
  author       = {Kuhl, Ellen},
  title        = {Continuum Mechanics},
  howpublished = {Lecture notes},
  note         = {me338, Stanford University},
  year         = {2003},
  url          = {https://biomechanics.stanford.edu/me338/kuhl_conti1.pdf},
}

@Article{Zhang2018,
author={Zhang, S. L.
and Wang, W. W.
and Burn, D. M.
and Peng, H.
and Berger, H.
and Bauer, A.
and Pfleiderer, C.
and van der Laan, G.
and Hesjedal, T.},
title={Manipulation of skyrmion motion by magnetic field gradients},
journal={Nature Communications},
year={2018},
month={May},
day={29},
volume={9},
number={1},
pages={2115},
issn={2041-1723},
doi={10.1038/s41467-018-04563-4},
url={https://doi.org/10.1038/s41467-018-04563-4}
}

@Article{Yang2024,
author={Yang, Yang
and Zhao, Le
and Yi, Di
and Xu, Teng
and Chai, Yahong
and Zhang, Chenye
and Jiang, Dingsong
and Ji, Yahui
and Hou, Dazhi
and Jiang, Wanjun
and Tang, Jianshi
and Yu, Pu
and Wu, Huaqiang
and Nan, Tianxiang},
title={Acoustic-driven magnetic skyrmion motion},
journal={Nature Communications},
year={2024},
month={Feb},
day={03},
volume={15},
number={1},
pages={1018},
issn={2041-1723},
doi={10.1038/s41467-024-45316-w},
url={https://doi.org/10.1038/s41467-024-45316-w}
}

@Article{Foerster2017,
author={Foerster, Michael
and Maci{\`a}, Ferran
and Statuto, Nahuel
and Finizio, Simone
and Hern{\'a}ndez-M{\'i}nguez, Alberto
and Lend{\'i}nez, Sergi
and Santos, Paulo V.
and Fontcuberta, Josep
and Hern{\`a}ndez, Joan Manel
and Kl{\"a}ui, Mathias
and Aballe, Lucia},
title={Direct imaging of delayed magneto-dynamic modes induced by surface acoustic waves},
journal={Nature Communications},
year={2017},
month={Sep},
day={01},
volume={8},
number={1},
pages={407},
issn={2041-1723},
doi={10.1038/s41467-017-00456-0},
url={https://doi.org/10.1038/s41467-017-00456-0}
}

@article{VANCAPEL201536,
title = {Nonlinear ultrafast acoustics at the nano scale},
journal = {Ultrasonics},
volume = {56},
pages = {36-51},
year = {2015},
issn = {0041-624X},
doi = {https://doi.org/10.1016/j.ultras.2014.09.021},
url = {https://www.sciencedirect.com/science/article/pii/S0041624X14002868},
author = {P.J.S. {van Capel} and E. Péronne and J.I. Dijkhuis}
}

@article{PhysRevB.86.134415,
  title = {Surface acoustic wave driven ferromagnetic resonance in nickel thin films: Theory and experiment},
  author = {Dreher, L. and Weiler, M. and Pernpeintner, M. and Huebl, H. and Gross, R. and Brandt, M. S. and Goennenwein, S. T. B.},
  journal = {Phys. Rev. B},
  volume = {86},
  issue = {13},
  pages = {134415},
  numpages = {13},
  year = {2012},
  month = {Oct},
  publisher = {American Physical Society},
  doi = {10.1103/PhysRevB.86.134415},
  url = {https://link.aps.org/doi/10.1103/PhysRevB.86.134415}
}

@Article{Birch2024,
author={Birch, Max T.
and Belopolski, Ilya
and Fujishiro, Yukako
and Kawamura, Minoru
and Kikkawa, Akiko
and Taguchi, Yasujiro
and Hirschberger, Max
and Nagaosa, Naoto
and Tokura, Yoshinori},
title={Dynamic transition and Galilean relativity of current-driven skyrmions},
journal={Nature},
year={2024},
month={Sep},
day={01},
volume={633},
number={8030},
pages={554-559},
issn={1476-4687},
doi={10.1038/s41586-024-07859-2},
url={https://doi.org/10.1038/s41586-024-07859-2}
}

@article{PhysRevB.108.054445,
  title = {Direct observations of spin fluctuations in hedgehog--anti-hedgehog spin lattice states in ${\mathrm{MnSi}}_{1\ensuremath{-}x}{\mathrm{Ge}}_{x} (x=0.6 \text{and} 0.8)$ at zero magnetic field},
  author = {Aji, Seno and Oda, Tatsuro and Fujishiro, Yukako and Kanazawa, Naoya and Saito, Hiraku and Endo, Hitoshi and Hino, Masahiro and Itoh, Shinichi and Arima, Taka-hisa and Tokura, Yoshinori and Nakajima, Taro},
  journal = {Phys. Rev. B},
  volume = {108},
  issue = {5},
  pages = {054445},
  numpages = {10},
  year = {2023},
  month = {Aug},
  publisher = {American Physical Society},
  doi = {10.1103/PhysRevB.108.054445},
  url = {https://link.aps.org/doi/10.1103/PhysRevB.108.054445}
}

\end{document}